\documentclass[5p,9pt,sort&compress]{elsarticle}

\usepackage{amsmath}
\usepackage{amssymb}
\usepackage{booktabs}
\usepackage{caption}
\usepackage{color}
\usepackage{float}
\usepackage{graphicx}
\usepackage[hidelinks]{hyperref}
\usepackage{multirow}
\usepackage{subcaption}
\usepackage{url}
\usepackage{wrapfig}
\usepackage{numprint}
\usepackage{tabularx}
\usepackage{geometry}
\usepackage{soulutf8}
\usepackage{comment}
\usepackage{xcolor}
\usepackage{colortbl}
\usepackage{apalike}
\usepackage{mathtools}

\graphicspath{{./figures_compressed/}}

\newcolumntype{Z}[0]{>{\centering\arraybackslash}X}%

\definecolor{artery}{HTML}{990099}
\definecolor{vein}{HTML}{006E6E}
\definecolor{rwnet}{HTML}{000764}
\definecolor{darkgreen}{HTML}{006400}
\definecolor{darkred}{HTML}{8B0000}
\definecolor{lightcyan}{HTML}{cfecff}
\definecolor{lightyellow}{HTML}{f7f6b5}

\soulregister\ref7
\soulregister\cite7
\soulregister\pageref7

\biboptions{authoryear}

\renewcommand{\cite}{\citep}

\makeatletter
\def\ps@pprintTitle{%
\def\@oddfoot{\footnotesize{Paper accepted for publication in \textit{Expert Systems with Applications}.}\hfill} 
}
\makeatother

\begin{document}

\title{%
    RRWNet: Recursive Refinement Network for effective retinal artery/vein segmentation and classification
}

\author[1]{José Morano\corref{cor1}}
\cortext[cor1]{Corresponding author.}
\ead{jose.moranosanchez@meduniwien.ac.at}

\author[1]{Guilherme Aresta}
\ead{guilherme.moreiraaresta@meduniwien.ac.at}

\author[1]{Hrvoje Bogunović}
\ead{hrvoje.bogunovic@meduniwien.ac.at }

\affiliation[1]{Christian Doppler Laboratory for Artificial Intelligence in Retina, Department of Ophthalmology and Optometry, Medical University of Vienna, Austria}

\date{}

\begin{abstract}

The caliber and configuration of retinal blood vessels serve as important biomarkers for various diseases and medical conditions.
A thorough analysis of the retinal vasculature requires the segmentation of the blood vessels and their classification into arteries and veins, typically performed on color fundus images obtained by retinography.
However, manually performing these tasks is labor-intensive and prone to human error.
While several automated methods have been proposed to address this task, the current state of art faces challenges due to manifest classification errors affecting the topological consistency of segmentation maps.
In this work, we introduce RRWNet, a novel end-to-end deep learning framework that addresses this limitation.
The framework consists of a fully convolutional neural network that recursively refines semantic segmentation maps, correcting manifest classification errors and thus improving topological consistency.
In particular, RRWNet is composed of two specialized subnetworks: a Base subnetwork that generates base segmentation maps from the input images, and a Recursive Refinement subnetwork that iteratively and recursively improves these maps.
Evaluation on three different public datasets demonstrates the state-of-the-art performance of the proposed method, yielding more topologically consistent segmentation maps with fewer manifest classification errors than existing approaches.
In addition, the Recursive Refinement module within RRWNet proves effective in post-processing segmentation maps from other methods,  further demonstrating its potential.
The model code, weights, and predictions will be publicly available at \url{https://github.com/j-morano/rrwnet}.

\end{abstract}


\maketitle

\section{Introduction}

\begin{figure}[tbp]
    \centering
    \includegraphics[width=0.48\textwidth]{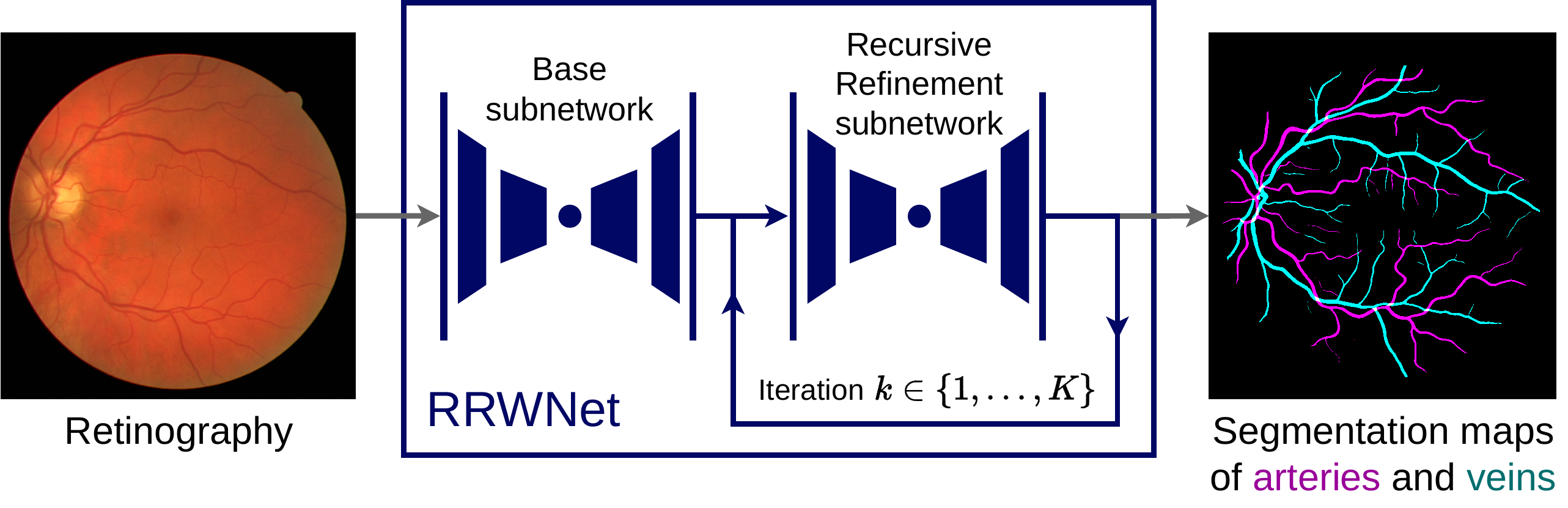}
    \caption{
        The proposed framework, \textcolor{rwnet}{\textbf{RRWNet}}, here applied for the segmentation and classification of retinal \textcolor{artery}{\textbf{arteries}} and \textcolor{vein}{\textbf{veins}},
        consists of a W-shaped fully convolutional neural network consisting of two subnetworks.
        The output of the first subnetwork (Base) is iteratively refined by the second (Recursive Refinement) through a recursive mechanism.
    }
    \label{fig:idea}
\end{figure}

The characteristics of retinal blood vessels (BV), including their caliber and configuration, serve as valuable biomarkers for diagnosing and monitoring several diseases and medical conditions, such as glaucoma, age-related macular degeneration (AMD), diabetic retinopathy (DR), and hypertension~\cite{Sun_SO_2009,Abramoff_RBE_2010,Kanski_Elsevier_2011}.
These alterations can be readily identified by trained ophthalmologists through analysis of color fundus images acquired via retinography, a non-invasive and cost-effective imaging technique that involves capturing photographs of the retina through the dilated pupil.
By virtue of its affordability and lack of invasiveness, retinography has become widely adopted in clinical practice, research investigations, and population-wide screening programs.
An example of a retinography image is shown in Fig.~\ref{fig:idea} (left).

A comprehensive analysis of the retinal vasculature requires the segmentation of blood vessels and their subsequent classification into arteries and veins (A/V).
This process yields separate A/V segmentation maps (as illustrated in Fig.~\ref{fig:idea}, right), enabling the quantification of various diagnostically relevant vessel characteristics, such as width, diameter, and tortuosity.
Furthermore, accurate measurement of these characteristics facilitates the calculation of more complex biomarkers, including the arteriolar-to-venular diameter ratio (AVR)~\cite{Ikram_IOVS_2004,Sun_SO_2009,Hatanaka_EMB_2005}.

However, manual execution of these tasks is inherently laborious, leading to increased costs, and is prone to human error, which negatively impacts both reproducibility and quality of care.
To address these limitations, several automated approaches have been proposed for the simultaneous segmentation and classification of arteries and veins~\cite{mookiah2021review}.

Current state-of-the-art methods predominantly leverage fully convolutional neural networks (FCNNs) \cite{Long_FullyConv_CVPR_2015} for this purpose~\cite{Galdran_Uncertainty_ISBI_2019,Hemelings_AV_CMIG_2019,Morano_AIIM_2021,Chen_MIA_2021,galdran2022sota,karlsson2022artery,hu2024semi}.
Most prevalent approaches classify each pixel into one of four classes: \emph{background}, \emph{artery}, \emph{vein}, and \emph{crossing} (representing regions where arteries and veins overlap).
Additionally, some methods incorporate an ``uncertain'' class to account for pixels presenting ambiguous characteristics~\cite{Galdran_Uncertainty_ISBI_2019,Hemelings_AV_CMIG_2019,Morano_AIIM_2021,Chen_MIA_2021,galdran2022sota,karlsson2022artery,hu2024semi}.
Conversely, some recent approaches~\cite{Chen_MIA_2021,Morano_AIIM_2021} frame the A/V segmentation and classification task as a multi-label segmentation problem.
This framework entails training the network to simultaneously segment arteries, veins, and BV (i.e., both arteries and veins) as separate classes, allowing a single pixel to be assigned to one or more classes.

\begin{figure}[t!]
    \centering
    \includegraphics[width=0.48\textwidth]{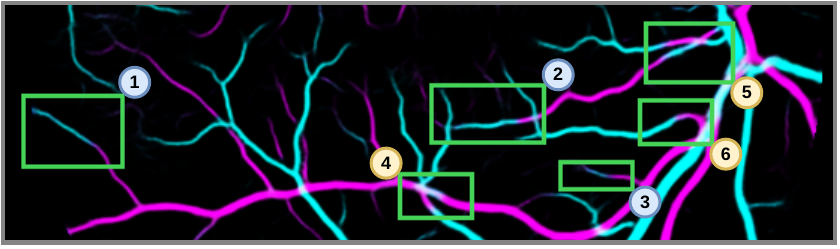}
    \caption{
        \sethlcolor{lightcyan}
        Examples of manifest classification errors produced by a state-of-the-art FCNN-based method~\cite{Morano_AIIM_2021}.
        \hl{\textbf{(1-3)}} While most of the vessel is classified as \textcolor{artery}{\textbf{artery}}, the model misclassifies the last part as \textcolor{vein}{\textbf{vein}}.
        \sethlcolor{lightyellow}
        \hl{\textbf{(4-6)}} The model often confuses the classification of vessels in crossing areas.
        These errors are easily detected by a human observer  because they are inconsistent with the overall structure of the vascular tree, hence the term ``manifest''.
    }
    \label{fig:manifest_errors}
\end{figure}

Irrespective of the chosen approach, state-of-the-art FCNN-based methods consistently encounter the challenge of \emph{manifest} classification errors.
These errors appear as regions where the predicted class contradicts the expected topological configuration of the target structures being segmented and classified.
In the context of A/V segmentation and classification, these errors translate to \emph{unreasonably} misclassified segments within predominantly correctly segmented vessels, as exemplified in Fig.~\ref{fig:manifest_errors}.
These errors arise from the propensity of FCNN-based models to classify vessels based on local characteristics of the input image, neglecting the global structural context of the vascular tree.
To mitigate these errors, several approaches have been proposed~\cite{Girard_Joint_AIM_2019,Kang_AV_CMPB_2020,Chen_MIA_2021,hu2024semi}.
Some methods~\cite{Girard_Joint_AIM_2019,Kang_AV_CMPB_2020} employ \emph{ad hoc} post-processing techniques based on graph propagation.
Alternatively, other methods~\cite{Chen_MIA_2021,hu2024semi} have proposed combining standard pixel-wise segmentation losses with adversarial losses~\cite{goodfellow2014generative} and custom-designed losses focused on specific characteristics of the predicted maps, such as topological consistency.
This combined approach aims to guide the model towards generating more topologically consistent segmentations.
Despite their contributions, these approaches exhibit limitations in their applicability beyond the specific task of blood vessel segmentation and classification (i.e., limited generalizability),
and their overall effectiveness in mitigating manifest classification errors remains limited.

In this work, we introduce RRWNet (Fig.~\ref{fig:idea}), a novel end-to-end deep learning framework for semantic segmentation that address the challenge of manifest classification errors in A/V segmentation and classification.
The proposed framework defines a recursive FCNN consisting of two specialized subnetworks: a Base subnetwork, which receives the input image and produces base segmentation maps, and a Recursive Refinement (RR) subnetwork, which receives the base segmentation maps and iteratively refines them, correcting manifest classification errors.
Extensive evaluation on multiple publicly available A/V segmentation and classification datasets demonstrates that the proposed RRWNet achieves state-of-the-art performance.
Specifically, it produces segmentation maps that exhibit superior classification accuracy and topological consistency, with a lower prevalence of manifest classification errors compared to existing methods.
Additionally, our experiments showcase the versatility of RRWNet by demonstrating the effectiveness of the RR subnetwork as a standalone post-processing technique.
This method significantly enhances the classification accuracy and topological consistency of segmentation maps produced by other methods.
For the sake of reproducibility, the source code, pre-trained model weights, and predicted results associated with RRWNet will be made publicly available on GitHub: \url{https://github.com/j-morano/rrwnet}.

\section{Related works}

\subsection{Vessel segmentation and classification}

The first methods for vessel segmentation on color fundus images were based on \textit{ad hoc} image processing techniques~\cite{Jiang_TPAMI_2003,Nain_MICCAI_2004,Tolias_TMI_1998,Staal_DRIVE_TMI_2004} or traditional learning models such as artificial neural networks~\cite{Sinthanayothin_BJO_1999,Marin_TMI_2011}.
Today, FCNNs based on the U-Net architecture~\cite{Ronneberger_U-Net_MICCAI_2015} have become the state of the art~\cite{Jiang_FullyConvVS_CMIG_2018,Oliveira_FullyConvVS_ESA_2018,Jin_DUNet_KBS_2019,wang2021csu,liu2022wave,liu2023aawgan,liu2023resdo}.

Until recently, A/V segmentation and classification was treated as a two-step process, where A/V classification was performed only on pixels previously identified as BV by a segmentation algorithm~\cite{Relan_EMBS_2014,Dashtbozorg_TIP_2014,Zamperini_Features_CBMS_2012,Estrada_TMI_2015,Welikala_CBM_2017}.
In addition, many works restricted the classification to small regions, usually around the optic disc~\cite{Zamperini_Features_CBMS_2012,Relan_EMBS_2014}.
The first work that deals with the whole vascular tree~\cite{Dashtbozorg_TIP_2014} proposed a graph-based method that takes as input a previously segmented vascular graph and the input image and obtains two separate graphs for arteries and veins.
Later, \citet{Welikala_CBM_2017} were the first to propose the use of a CNN for the classification stage.
Although these methods achieved reasonable performance, they were limited by the quality of the initial vessel segmentation.

To avoid this problem, several works have addressed the simultaneous segmentation and classification of retinal vessels either as a semantic segmentation task of three to four classes (background, artery, vein, uncertain)~\cite{Xu_AV_BOE_2018,Girard_Joint_AIM_2019, Hemelings_AV_CMIG_2019,Galdran_Uncertainty_ISBI_2019,Ma_MultiTask_MICCAI_2019,Kang_AV_CMPB_2020,Morano_AIIM_2021,Chen_MIA_2021,karlsson2022artery,galdran2022sota} or as a multi-label segmentation task with multiple targets~\cite{Morano_AIIM_2021,Chen_MIA_2021} (arteries, veins, blood vessels).
The latter approach has the advantage of providing continuous and thus more topologically consistent segmentation maps of arteries and veins, since vessel crossings are considered as both classes at the same time.

\subsection{Reducing manifest classification errors}

Most of the aforementioned studies acknowledge the challenge of manifest classification errors, resulting from models favoring local input characteristics over the global structure of the vascular tree.
Existing approaches to address this problem can be broadly categorized into two groups: \emph{ad hoc post-processing} and \emph{learning based}.

Ad hoc post-processing methods~\cite{Girard_Joint_AIM_2019,Kang_AV_CMPB_2020} typically involve graph-based operations on the vessel graph extracted from previously segmented maps.
These methods usually require additional input information (such as the location of the optic disc), and their effectiveness is severely limited by the quality of the initial segmentation maps, which hinders their generalization capabilities.

Conversely, learning-based methods~\cite{hu2024semi,Chen_MIA_2021,karlsson2022artery} aim to address manifest classification errors by incorporating additional losses and architectural modifications.
\citet{hu2024semi} introduced a multi-class point consistency module to generate artery and vein skeletons, which are then used to compute different consistency losses aimed at improving topological consistency and mitigating classification errors. 
Similarly, \citet{Chen_MIA_2021} proposed a GAN-based method with a topological loss.
In particular, they proposed to use a discriminator designed to rank, from lowest to highest, the topological connectivity of the ground truth, the predicted mask, and a randomly transformed mask.
The ranking error is used as a loss to encourage the model to produce more topologically consistent segmentation maps.
In addition, they proposed a new module that extracts the high-level topological features of the images to force the model to predict vascular segmentation maps with a topology similar to that of the manual annotations.
Although these methods are interesting and fairly effective,
they rely on task-specific mechanisms, and their performance remains limited.

In contrast to these methods, we rely on a recursive refinement approach with two specialized subnetworks that implicitly leverage local and global information to iteratively correct manifest classification errors.

\subsection{Iterative refinement}

In recent years, several iterative refinement approaches have been proposed to improve segmentation performance in both natural and medical image analysis~\cite{Pinheiro_ICML_2014,Sironi_TPAMI_2016,Newell_ECCV_2016,Shen_ICCV_2017,Januszewski_arXiv_2016,Mosinska_CVPR_2018,karlsson2022artery,galdran2022sota}.
These methods use an iterative prediction process via a classifier that receives as input the result from the previous iteration(s) and, optionally, the input image, addressing the errors made in earlier iterations.

A common approach consists of \emph{stacking} multiple deep modules and training them in an end-to-end fashion~\cite{Newell_ECCV_2016,Shen_ICCV_2017,karlsson2022artery,galdran2022sota}.
This allows for modules specialized in solving the errors of the previous modules.
For example, \citet{Newell_ECCV_2016} proposed a novel architecture composed of eight consecutive modules for pose estimation in natural images.
During training, all modules receive supervision through comparison of their outputs with the ground truth.
However, such methods often require a substantial number of parameters, leading to high memory and computational costs during both training and inference.
To address this limitation, recent work has proposed the use of very lightweight encoder-decoder networks~\cite{karlsson2022artery,galdran2022sota}.
Using a custom architecture consisting of four U-Net-like networks, 
\citet{karlsson2022artery} achieved state-of-the-art performance on various A/V segmentation and classification benchmarks.
However, their approach requires extensive hyperparameter tuning to perform well, and the optimal hyperparameters were found to be dataset-specific.
Their final model was achieved through an exhaustive search exploring various network configurations and loss functions.
In particular, the authors experimented with different numbers of networks, layers, levels, and kernels in each network/layer, as well as different weights for the loss terms and the regularization parameters.
Additionally, the generalization capabilities of the method have not been thoroughly evaluated.
\citet{galdran2022sota} proposed a similar approach for the same task using only two stacked U-Net-like networks, although with limited performance.

An alternative approach consists of refining the predictions using a single \emph{recursive} network~\cite{Pinheiro_ICML_2014,Mosinska_CVPR_2018}.
While the memory requirements during training of this approach remain constant, as the gradients for each iteration must be stored, the number of parameters is much lower, which makes it more efficient at test time.
In this line of work, \citet{Pinheiro_ICML_2014} proposed to perform semantic segmentation of natural images by using a CNN network that subsequently refines the predicted maps at different scales.
This approach is focused on increasing the spatial context of the network, so that it models non-local dependencies (of higher level) in the scenes.
In this way, the authors manage to make the network give rise to more structurally coherent predictions.
The problem with this method is that the CNN is applied \textit{pixel-wise}, making it very inefficient in terms of computational cost.
This problem is addressed by \citet{Mosinska_CVPR_2018} by using a FCNN at a constant scale.
Their method involves training a FCNN that recursively refines an initial segmentation over multiple iterations.
In the first iteration, the network receives the input image and an empty segmentation map (all zeros).
In subsequent iterations $k \in \{1,...,K\}$, it receives the same image together with the segmentation map obtained in iteration $k-1$.
The training loss function used to train the network is a weighted sum of the losses from all the iterations, with higher weights assigned to later iterations.
Despite promising results, this approach exhibits lower performance compared to stacking modules.
This may be attributed to the lack of specialized refinement modules, since the same network needs to leverage both relatively local information for initial segmentation and more structural and global information for further refinement. 
This limits the ability of the model to address the specific challenges of progressive segmentation refinement.

Leveraging the strengths of both stacking and recursive approaches, our framework innovatively decomposes an FCNN into two specialized parts: a Base subnetwork, which generates a base segmentation,
and a RR subnetwork, which iteratively refines the base segmentation with the goal of resolving manifest classification errors.

\section{Contributions}

The main contributions of our work are as follows:

\begin{enumerate}
    \item We propose RRWNet, a novel end-to-end deep learning framework for recursively refining semantic segmentation maps to correct manifest classification errors.
    Our framework is the first to combine the advantages of module stacking and recursive refinement approaches by decomposing the network into two specialized parts, a Base subnetwork and a Recursive Refinement subnetwork, which are trained jointly in an end-to-end manner.
    \item We propose and publicly release a straightforward implementation of the proposed framework, based on FCNNs, for the automatic segmentation and classification of retinal vessels into arteries and veins in retinography images.
    \item We demonstrate that RRWNet achieves state-of-the-art performance in A/V segmentation and classification on various public datasets (RITE, LES-AV, and HRF), showcasing the effectiveness of our framework.
    \item Furthermore, we show that Recursive Refinement subnetwork of RRWNet can be used as an effective standalone post-processing technique, significantly improving the classification accuracy and the topological consistency of segmentation maps generated by other state-of-the-art methods.
\end{enumerate}

\section{Methods}

\begin{figure*}
    \centering
    \includegraphics[width=0.8\textwidth]{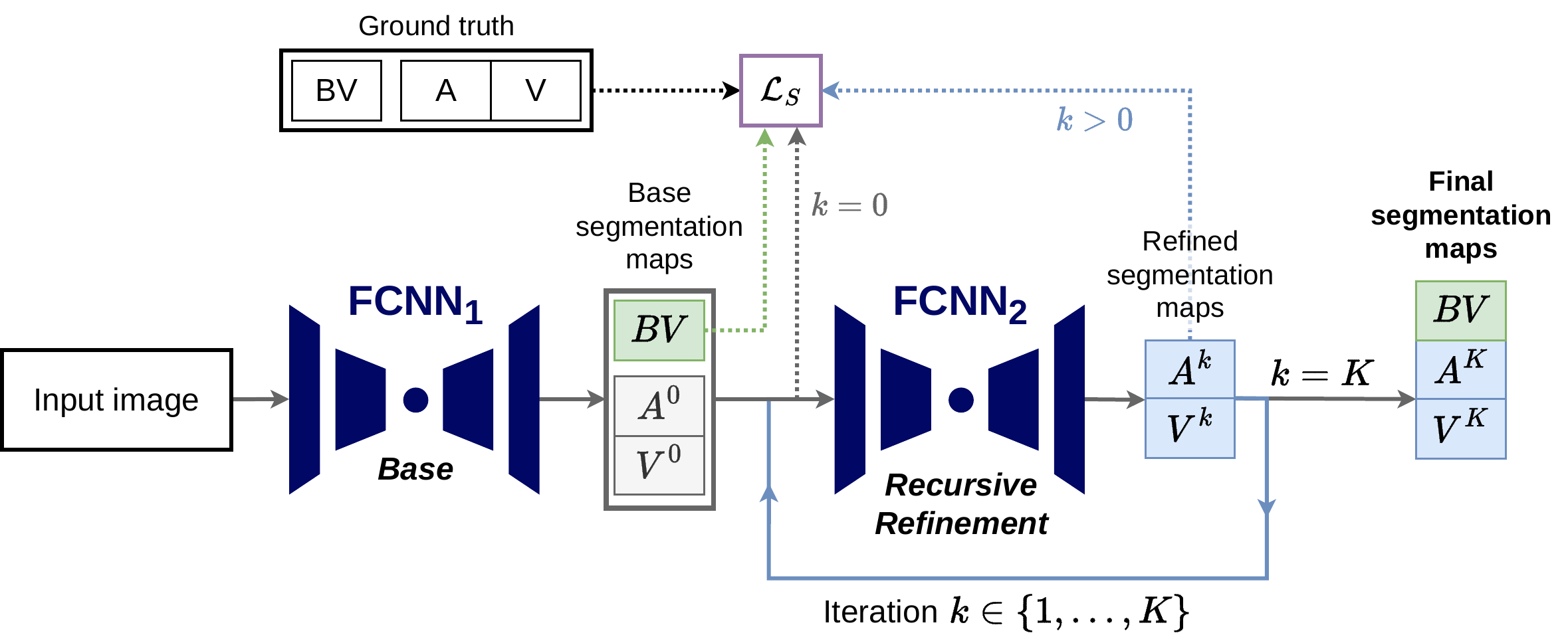}
    \caption{
        Proposed approach for the segmentation and classification of arteries and veins.
        The input image is fed to the Base subnetwork, which produces coarse segmentation maps of arteries (A), veins (V) and blood vessels (BV).
        Next, the A/V segmentation maps are fed to the Recursive Refinement subnetwork, which recursively refines them for a certain number of iterations $K$.
    }
    \label{fig:approach}
\end{figure*}

Fig.~\ref{fig:approach} provides a detailed view of the proposed RRWNet framework, focusing on its application to A/V segmentation and classification.
The Base subnetwork takes as input a retinography image and produces base segmentation maps of arteries (A), veins (V), and blood vessels (BV) (the union of arteries and veins).
These segmentation maps (without the input image) are then fed to the RR subnetwork.
This subnetwork recursively refines the segmentation maps of arteries and veins a certain number of iterations $K$, iteratively correcting manifest vessel classification errors made by the Base subnetwork.
BV segmentation is not refined based on previous work indicating high accuracy with a single U-Net \cite{Morano_AIIM_2021,karlsson2022artery}.
In each iteration, the input of the RR subnetwork is exclusively the output of the preceding iteration.
This forces the network to focus on correcting errors based on the existing blood vessel structure (whose segmentation remains fixed through the different iterations), rather than relying on the characteristics of the input image.
The final output of the network consists of the refined A/V segmentation maps at the last iteration ($k=K$) and the initial BV segmentation map produced by the Base subnetwork.

Specifically, let $\mathbf{x} \in \mathbb{R}^{3 \times H \times W}$ be the input RGB retinography image, where $H$ and $W$ are its height and width, respectively.
Similarly, the ground truth (GT) $\mathbf{y} \in \mathbb{R}^{3 \times H \times W}$ also has three channels, corresponding to the manual segmentation maps of arteries ($\mathbf{y}^A$), veins ($\mathbf{y}^V$), and vessels ($\mathbf{y}^{BV}$). 
We obtain the final prediction $\mathbf{\hat{y}} \in \mathbb{R}^{3 \times H \times W}$.
This prediction is obtained by applying the network $f(\mathbf{x}, \theta, K)$ to the input image $\mathbf{x}$, where $\theta$ represents the learnable parameters of the network and $K$ denotes the number of iterations performed by the RR subnetwork.
Thus, the output of the network $\mathbf{\hat{y}}_k$ at an arbitrary iteration $k$ can be defined as
\begin{equation}\label{eq:network}
    \small
    \begin{aligned}
        \mathbf{\hat{y}}_k &= f(\mathbf{x}, \theta, k) \\
        &= \begin{cases}
            f_R(f(\mathbf{x}, \theta, k-1), \theta_R)^{A,V} \oplus f_B(\mathbf{x}, \theta_B)^{BV}, & k > 0 \\
            f_B(\mathbf{x}, \theta_B), & k = 0
        \end{cases}
    \end{aligned}
\end{equation}
where
$f_B$ and $f_R$ represent the Base and RR subnetworks with parameters $\theta_B$ and $\theta_R$, respectively,
the superscripts $A$, $V$, and $BV$ denote the channels corresponding to the segmentation maps of the different structures,
and $\oplus$ denotes the concatenation operation in the channel dimension.

\paragraph{Training loss}
To train the network, we use a loss function $\mathcal{L}$ that combines the segmentation errors from each iteration $k \in \{0,...,K\}$ with different weights.
This loss function is defined as:
\begin{equation}\label{eq:loss}
    \mathcal{L}\left(\mathbf{\hat{y}},\mathbf{y}\right) ~=~ \sum_{k=0}^{K} w_k \mathcal{L}_\text{S}\left(\mathbf{\hat{y}}_k,\mathbf{y}\right) \ ,
\end{equation}
where
$w_k$ is the weight of the loss at iteration $k$,
and $\mathcal{L}_\text{S}$ is the segmentation loss function, defined as the sum of individual binary segmentation errors for each structure (arteries, veins, and BV).
Following previous works~\cite{Morano_AIIM_2021,Chen_MIA_2021}, we use the Binary Cross-Entropy (BCE) loss between the prediction and the GT for each structure.
Thus, the segmentation loss $\mathcal{L}_\text{S}$ for $N$ structures is defined as:
\begin{equation}\label{eq:loss_S}
    \mathcal{L}_\text{S}\left(\mathbf{\hat{y}},\mathbf{y}\right) ~=~ \sum_{i=1}^{N}\text{BCE}\left(\mathbf{\hat{y}}_i,\mathbf{y}_i\right) \ ,
\end{equation}
where $\mathbf{\hat{y}}_i$ and $\mathbf{y}_i$ are the $i$-th channel of the output of the network and the GT, respectively, representing individual structures.
The weighting scheme is similar to that of \citet{mookiah2021review}, with the difference that we give a higher weight to the error of the first iteration ($k=0$), which solely relies on the Base subnetwork.
This prioritizes producing fairly accurate base segmentations before applying subsequent refinements.
Specifically, the weight $w_k$ at iteration $k$ is defined as
\begin{equation}\label{eq:weights}
    w_k =
    \begin{cases}
        1, & k = 0 \\
        \frac{1}{Z} \sum_{k=1}^{K} k \mathcal{L}_\text{S}\left(\mathbf{\hat{y}}_k,\mathbf{y}_k\right), & k > 0
    \end{cases}
\end{equation}
with the normalization factor $Z = \sum_{k=1}^{K} k$. 

\paragraph{Network architecture}\label{sec:network}

\begin{figure}
    \centering
    \includegraphics[width=0.48\textwidth]{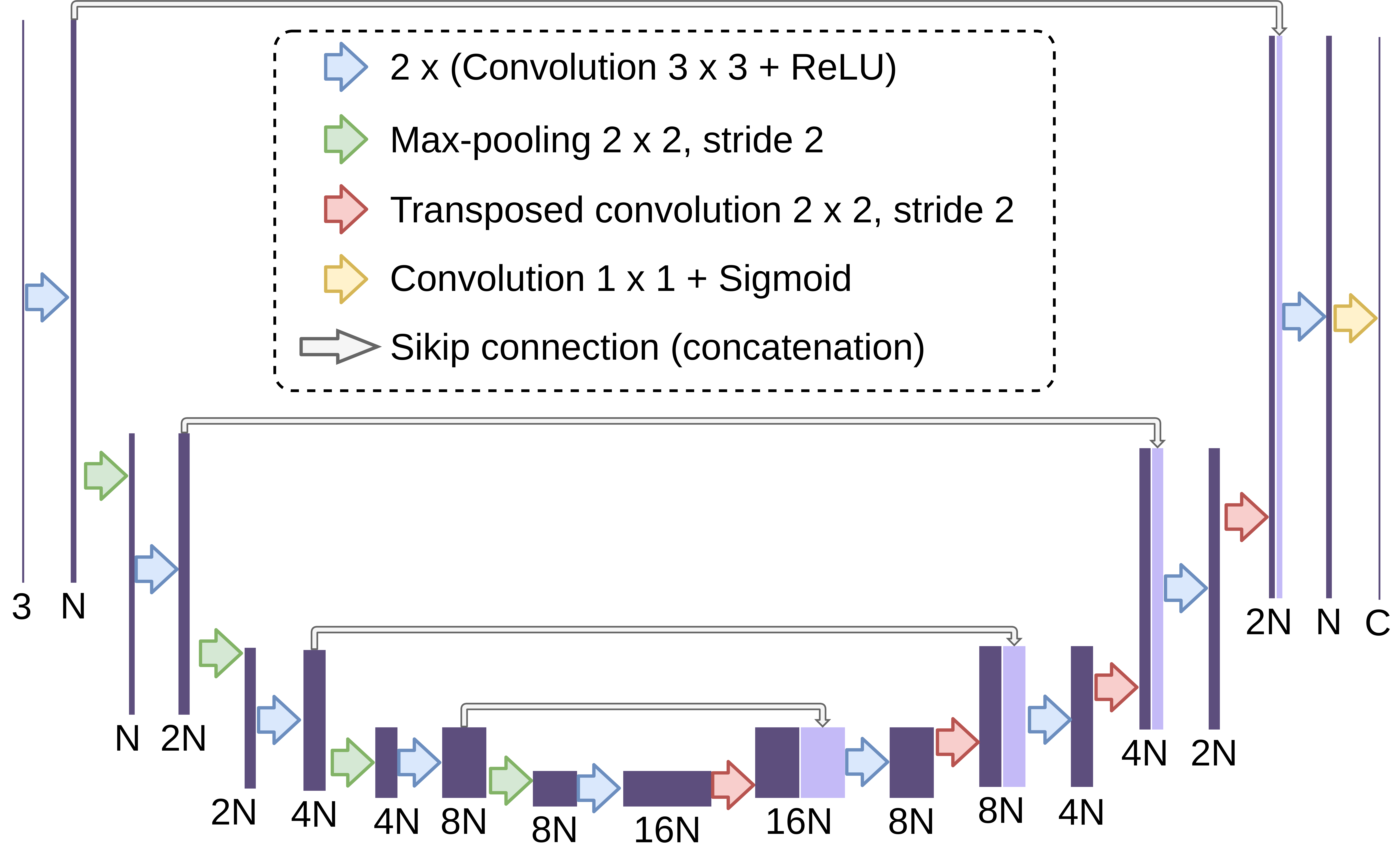}
    \caption{
        U-Net architecture of the two subnetworks.
        $N$ represents the number of base channels.
        $C$ represents the number of output channels.
        In our case, $N=64$.
    }
    \label{fig:U-Net}
\end{figure}

The proposed network architecture consists of two nearly-identical encoder-decoder subnetworks connected in series ($\text{FCNN}_1$ and $\text{FCNN}_2$, in Fig.~\ref{fig:approach}).
While architecturally similar, the subnetworks differ in their function and the number of output channels.
The first subnetwork, used to obtain the base segmentation maps, has 3 output channels (arteries, veins, and BV), and the second, used for iterative refinement, has 2 (arteries and veins).
Similarly to the state of the art~\cite{Xu_AV_BOE_2018,Girard_Joint_AIM_2019,Hemelings_AV_CMIG_2019,Galdran_Uncertainty_ISBI_2019,Ma_MultiTask_MICCAI_2019,Morano_ECAI_2020,Morano_AIIM_2021}, we adopt the original U-Net architecture \cite{Ronneberger_U-Net_MICCAI_2015} for both subnetworks.
At the end of each subnetwork, a sigmoid function is used to produce the segmentation maps of all structures.
A complete diagram of the U-Net architecture used in this work is shown in Fig.~\ref{fig:U-Net}.

\section{Experimental setup}

\subsection{Datasets}\label{subsec:data}

\begin{figure*}
    \subfloat[RITE retinography.]{\includegraphics[width=0.33\linewidth]{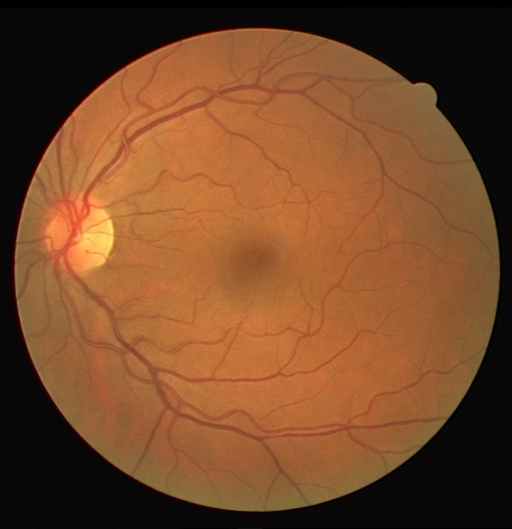}}
    \hfill
    \subfloat[RITE~\cite{Hu_RITE_MICCAI_2013} annotations.*]{\includegraphics[width=0.33\linewidth]{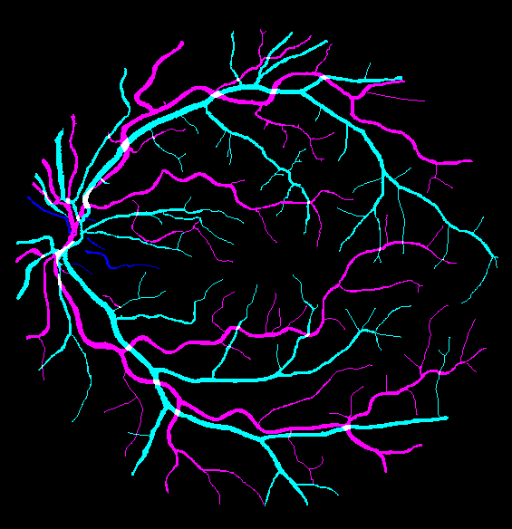}}
    \hfill
    \subfloat[\citet{Qureshi_ISCBMS_2013} annotations.\textsuperscript{\textdagger}]{\includegraphics[width=0.33\linewidth]{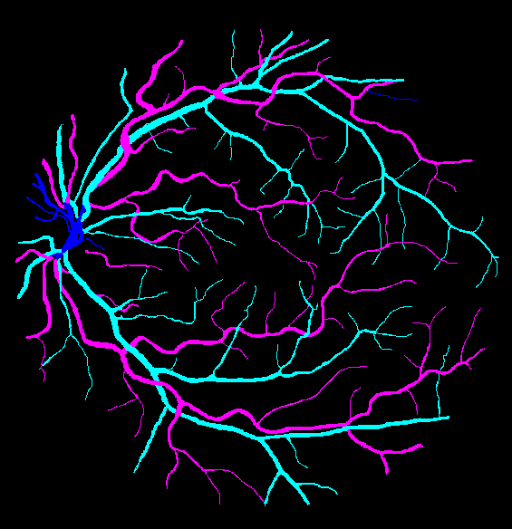}}

    \vspace{0.3cm}

    \subfloat[LES-AV retinography.]{\includegraphics[width=0.33\linewidth]{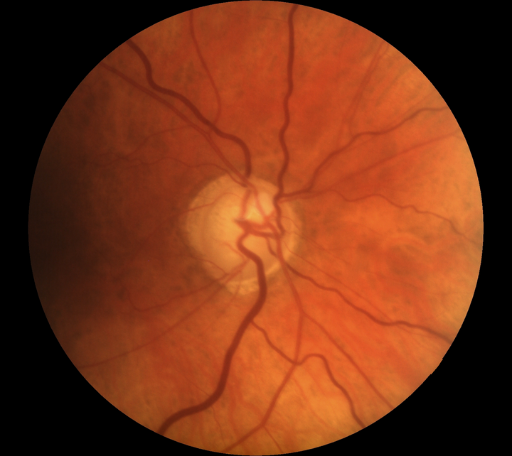}}
    \hspace{0.3cm}
    \subfloat[\citet{Orlando_LESAV_MICCAI_2018} annotations.*]{\includegraphics[width=0.33\linewidth]{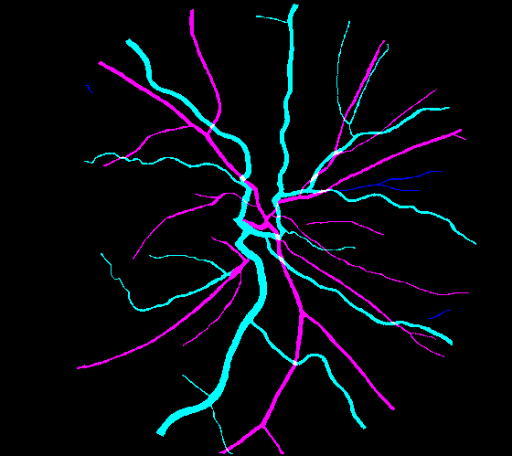}}

    \vspace{0.3cm}

    \subfloat[HRF retinography.]{\includegraphics[width=0.33\linewidth]{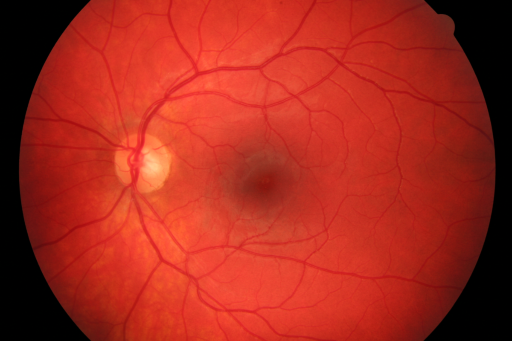}}
    \hfill
    \subfloat[\citet{Chen_MIA_2021} annotations.*]{\includegraphics[width=0.33\linewidth]{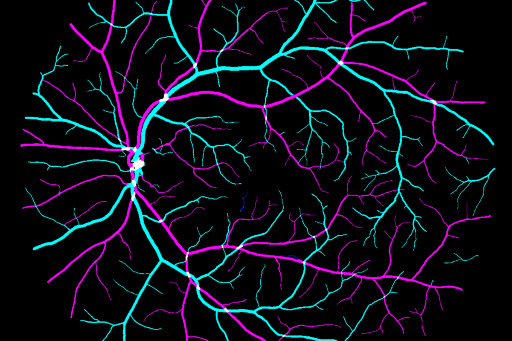}}
    \hfill
    \subfloat[\citet{Hemelings_AV_CMIG_2019} annotations.\textsuperscript{\textdagger}]{\includegraphics[width=0.33\linewidth]{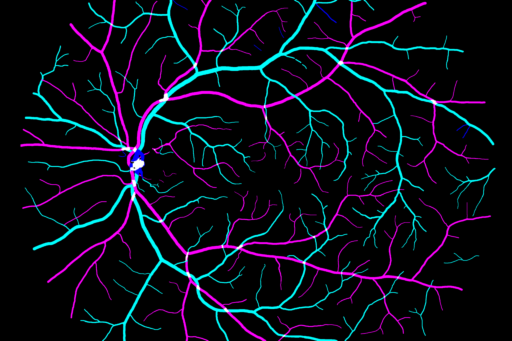}}
    \caption{
        Examples of retinography images from different datasets
        and their corresponding A/V segmentation maps.
        (a-d) RITE.
        (e,f) LES-AV.
        (f-h) HRF.
        The segmentation maps are visualized as RGB images composed of the segmentation maps of \textcolor{artery}{\textbf{arteries}} (Red channel),
        \textcolor{vein}{\textbf{veins}} (Green channel) and \textbf{vessels} (Blue channel).
        This composition makes arteries appear magenta, veins appear cyan, crossings appear white (because they are both arteries and veins at the same time), and uncertain vessels appear blue (because they are not assigned to either artery or vein class, but only to vessel).
        *~Annotations used for training and testing.
        \textdagger~Second expert annotations.
    }
    \label{fig:examples_datasets}
\end{figure*}

Experiments were performed on the three publicly available datasets containing color fundus images with corresponding A/V annotations:
RITE~\cite{Hu_RITE_MICCAI_2013}, LES-AV~\cite{Orlando_LESAV_MICCAI_2018}, and HRF~\cite{budai2013robust}.
Fig.~\ref{fig:examples_datasets} shows some examples of color fundus images and their
corresponding GT segmentation maps from the three datasets, while Table~\ref{tab:sample_distribution} provides an overview of the distribution of samples (pixels) across the different classes (background, artery, vein, crossing, uncertain) within each dataset.
Further details on the datasets are provided below.

\paragraph{Retinal Images vessel Tree Extraction (RITE)}
The RITE\footnote{\url{https://medicine.uiowa.edu/eye/rite-dataset} (accessed on 2023-12-15)} dataset \cite{Hu_RITE_MICCAI_2013} (also known as AV-DRIVE or DRIVE-AV in the literature) is an extension of the Digital Retinal Images for Vessel Extraction (DRIVE) dataset~\cite{Staal_DRIVE_TMI_2004}.
While DRIVE focuses on vessel segmentation, RITE incorporates additional manual GT segmentation maps for classifying arteries and veins.
The dataset comprises the same 40 retinography of DRIVE (20 for training and 20 for testing).
These images originate from 33 healthy patients and 7 patients with mild signs of DR.
They are all centered on the macula and have a resolution of $768 \times 584$ pixels, with a circular region of interest (ROI).
RITE includes the original blood vessel segmentation maps from DRIVE, along with the pixel-level classification of these vessels into arteries, veins, crossings, and uncertain classes~\cite{Hu_RITE_MICCAI_2013}.
Crossings indicate areas where a vein and an artery overlap.
The ``uncertain'' class is used for those vessels whose classification the experts have not been able to determine.
Alternative A/V classification annotations for the DRIVE dataset were proposed by \citet{Qureshi_ISCBMS_2013}.
They manually segmented and classified the blood vessels into arteries, veins, and uncertain from the raw retinography images, independent of the existing DRIVE vessel segmentations.
For artery-vein crossings, the classification was assigned based on the vessel closest to the surface of the retina.
These alternative labels, henceforth referred to as Qureshi et al., were used in this work as additional ``second expert'' annotations.
\begin{table}
    \newcommand{\tabitem}{~~~-~}
    \centering
    \caption{%
        Distribution of samples (pixels) among the different classes in the different datasets for the labels used for training and evaluation.
        All values are percentages.
    }
    \label{tab:sample_distribution}
    \begin{tabular}{@{\extracolsep{4pt}}lrrr}
    \toprule
        \textbf{Class} & \multicolumn{3}{c}{\textbf{Dataset}} \\
        \cmidrule{2-4}
        & RITE
        & LES-AV
        & HRF
        \\

        \midrule

        Background & 87.52 & 90.50 & 89.88 \\
        \midrule
        Vessel & 12.48 & 9.50 & 10.12 \\
        \tabitem \textcolor{artery}{Artery} & 5.19 & 4.28 & 4.49 \\
        \tabitem \textcolor{vein}{Vein} & 6.37 & 4.81 & 5.19 \\
        \tabitem Crossing & 0.32 & 0.14 & 0.26 \\
        \tabitem \textcolor{blue}{Uncertain} & 0.60 & 0.27 & 0.18 \\
    \bottomrule
    \end{tabular}%
\end{table}

\paragraph{LES-AV}
This dataset\footnote{\url{https://figshare.com/articles/dataset/LES-AV_dataset/11857698} (accessed on 2023-12-16)}~\cite{Orlando_LESAV_MICCAI_2018}
consists of 22 retinography images, originating from both healthy patients (11) and patients with signs of glaucoma (11).
Unlike RITE, LES-AV does not have a predefined split for training and testing.
The images are centered on the optic disc and have a resolution of $1620 \times 1444$ pixels (except one image with a resolution of $2196 \times 1958$ pixels) and a circular ROI.
Similarly to RITE, LES-AV includes GT segmentation maps for blood vessels, classified into arteries, veins, crossings, and uncertain regions.
In this work, we employ LES-AV as an external dataset for cross-dataset evaluations to assess the generalization capabilities of the proposed method.

\paragraph{High-Resolution Fundus (HRF)}
The HRF dataset\footnote{\url{https://www5.cs.fau.de/research/data/fundus-images/} (accessed on 2023-12-16)}~\cite{budai2013robust} is a collection of 45 high-resolution retinography images ($3504 \times 2336$ pixels).
The images are categorized into three groups: 15 images from healthy individuals, 15 from patients with diabetic retinopathy (DR), and 15 from glaucomatous patients.
A/V classification annotations were provided by two different sources.
\citet{Hemelings_AV_CMIG_2019}\footnote{\url{https://github.com/rubenhx/av-segmentation} (accessed on 2023-12-16)} provided the original annotations used for the dataset.
Then, \citet{Chen_MIA_2021}\footnote{\url{https://github.com/o0t1ng0o/TW-GAN} (accessed on 2023-12-16)} introduced novel manual annotations to address inconsistencies in the annotations of \citet{Hemelings_AV_CMIG_2019}.
The annotation procedures of these two works differ slightly in handling artery-vein crossings.
While \citet{Chen_MIA_2021} label crossings as a separate class (consistent with RITE and LES-AV), \citet{Hemelings_AV_CMIG_2019} assign them to the uppermost vessel if known, or to the uncertain class otherwise.
In this work, we primarily use the \citet{Chen_MIA_2021} annotations for training and testing, while we use the \citet{Hemelings_AV_CMIG_2019} annotations as ``second expert'' annotations.
Following previous works~\cite{Hemelings_AV_CMIG_2019,karlsson2022artery}, we use the first five images in each category for testing and the remainder for training.

\subsection{Experiments}

\paragraph{Hyperparameter search}
The number of refinement steps $K$ is an important hyperparameter of RRWNet.
For this reason, we conducted a grid search on the RITE dataset (validation) to determine its optimal value.
In particular, we evaluated the performance of RRWNet with the following values of $K$: 2, 3, 6, 8, and 11.
The evaluation was based on the average of the AUROC and AUPR for artery, vein, and BV segmentation, as well as the accuracy of A/V and BV/BG classification.
The value that yielded the best overall performance according to these metrics was selected for subsequent experiments.

\paragraph{Ablation study}
We performed an ablation study to evaluate the individual and combined impact of the proposed RR module and the recursive refinement (multiple RR applications).
For this end, we compared the following models:
    (1) \emph{U-Net}: A standard U-Net architecture \cite{Ronneberger_U-Net_MICCAI_2015}. This is equivalent to use only the Base subnetwork of RRWNet.
    (2) \emph{W-Net}: Our proposed approach without recursive refinement (i.e., $K=1$). This isolates the effect of the recursive refinement.
    (3) \emph{RRU-Net}: An architecture similar to \citet{Mosinska_CVPR_2018}, that employs recursive refinement with a single encoder-decoder. This is equivalent to use only the RR subnetwork of RRWNet in a recursive manner while providing the input image.
    (4) \emph{RRWNetAll}: Our full model (Base and RR subnetworks) refining all segmentation maps (arteries, veins, and BV).
    (5) \emph{RRWNet}: Our full model refining only the A/V segmentation maps.
All experiments were conducted on the RITE dataset. Statistical significance was assessed using the one-tailed Wilcoxon signed-rank test.

\paragraph{State-of-the-art comparison and recursive refinement post-processing}
We compared RRWNet with several state-of-the-art methods for A/V segmentation and classification on RITE, LES-AV and HRF datasets: \citet{Girard_Joint_AIM_2019,Galdran_Uncertainty_ISBI_2019,Ma_MultiTask_MICCAI_2019,Hemelings_AV_CMIG_2019,Kang_AV_CMPB_2020,Morano_AIIM_2021,Chen_MIA_2021,galdran2022sota,hatamizadeh2022ravir,karlsson2022artery,hu2024semi}.
To comprehensively evaluate our approach, we used different evaluation protocols.
In the first place, we employed the standard evaluation protocol in the field, focusing on the A/V classification accuracy for the intersection of predicted and GT vessels.
Additionally, we 
performed a more in depth comparison
with recent state-of-the-art methods that provided source code or continuous predicted segmentation maps for any of the datasets.
In particular, following \citet{Morano_AIIM_2021} and \citet{Chen_MIA_2021}, we compared the methods in terms of A/V classification performance for all vessel pixels in the GT,  as well as as A/V segmentation performance using and several threshold-independent and topological metrics.
The methods included in the comparison were
\citet{Morano_AIIM_2021}, \citet{Chen_MIA_2021}\footnote{\url{https://github.com/o0t1ng0o/TW-GAN} (accessed on 2023-12-16)},
\citet{karlsson2022artery}\footnote{\url{https://github.com/robert-karlsson/av-segmentation} (accessed on 2023-12-16)},
and \citet{galdran2022sota}~\footnote{\url{https://github.com/agaldran/lwnet} (accessed on 2023-12-16)}.
In order to standardize the evaluation criteria, we did not perform any post-processing on the segmentation maps produced by these methods.
Following previous work~\cite{Galdran_Uncertainty_ISBI_2019,Chen_MIA_2021,galdran2022sota}, models were trained and tested separately on RITE and HRF, while LES-AV was used for cross-dataset evaluation (trained on RITE, tested on LES-AV).

Finally, to assess the generalizability and potential of the RR subnetwork as a post-processing technique, we evaluated its performance when applied to the segmentation maps generated by the aforementioned state-of-the-art methods.

\begin{table*}[t!]
    \centering
    \newcommand{\cmidr}{\cmidrule{2-8}}
    \caption{Impact of $K$ in RITE (validation).
    Proposed RRWNet with different $K$: 2, 3, 6, 8, and 11.
    Best results are highlighted in \textbf{bold}, and second best results are \underline{underlined}.
    All values are percentages.
    }
    \label{tab:search_k}
    \resizebox{\linewidth}{!}{%
    \begin{tabular}{@{\extracolsep{4pt}}lllccccc}
    \toprule

    \textbf{Evaluation}
    & \textbf{Structure}
    & \textbf{Metric}
    & \multicolumn{5}{c}{\textbf{Models}}
    \\

    \cmidrule{4-8}

    & &
    & \multicolumn{1}{c}{RRWNet-2}
    & \multicolumn{1}{c}{RRWNet-3}
    & \multicolumn{1}{c}{\textit{RRWNet-6}}
    & \multicolumn{1}{c}{RRWNet-8}
    & \multicolumn{1}{c}{RRWnet-11}
    \\

    \midrule

    \multirow{6}{*}{Segmentation}
    & \multirow{2}{*}{Artery}
    & AUROC
    & $98.19$
    & $98.50$
    & \underline{$98.55$}
    & $98.53$
    & $\mathbf{98.59}$
    \\

    &
    & AUPR
    & $84.93$
    & $85.96$
    & $86.08$
    & \underline{$86.14$}
    & $\mathbf{86.20}$
    \\

    \cmidr

    & \multirow{2}{*}{Vein}
    & AUROC
    & $98.47$
    & $98.66$
    & $\mathbf{98.75}$
    & $98.72$
    & \underline{$98.73$}
    \\

    &
    & AUPR
    & $89.22$
    & \underline{$89.42$}
    & $\mathbf{89.64}$
    & $89.35$
    & $89.39$
    \\

    \cmidr

    & \multirow{2}{*}{BV}
    & AUROC
    & $97.98$
    & \underline{$98.03$}
    & $97.98$
    & $\mathbf{98.06}$
    & \underline{$98.03$}
    \\

    &
    & AUPR
    & $90.83$
    & $90.93$
    & $90.98$
    & \underline{$91.02$}
    & $\mathbf{91.04}$
    \\

    \midrule

    \multirow{6}{*}{Classification}
    & \multirow{3}{*}{Artery/Vein}
    & Sens.
    & $94.70$
    & $\mathbf{95.31}$
    & \underline{$95.13$}
    & $94.89$
    & $94.89$
    \\

    &
    & Spec.
    & $95.39$
    & $\mathbf{96.16}$
    & \underline{$96.00$}
    & $95.88$
    & $95.72$
    \\

    &
    & Acc.
    & $95.08$
    & $\mathbf{95.78}$
    & \underline{$95.61$}
    & $95.43$
    & $95.35$
    \\

    \cmidr

    & \multirow{3}{*}{BV/BG}
    & Sens.
    & $\mathbf{79.20}$
    & \underline{$79.12$}
    & $77.80$
    & $78.29$
    & $78.49$
    \\

    &
    & Spec.
    & $97.99$
    & $98.03$
    & $\mathbf{98.25}$
    & \underline{$98.16$}
    & \underline{$98.16$}
    \\

    &
    & Acc.
    & $95.64$
    & $95.67$
    & \underline{$95.69$}
    & $95.68$
    & $\mathbf{95.70}$
    \\

    \bottomrule
    \end{tabular}%
    }
\end{table*}

\subsection{Evaluation metrics}\label{sec:metrics}

Segmentation performance was evaluated using receiver operating characteristic (ROC) curves, precision-recall (PR) curves, and one-versus-all classification metrics (sensitivity, specificity, and accuracy) for each structure of interest.
We calculated the area under the curve (AUC) value to summarize the information from the curves.
The use of PR curves along with ROC curves was motivated by their greater sensitivity to imbalanced classes \cite{davis2006relationship,saito2015precision}, as encountered here with arteries, veins, and background (see Table~\ref{tab:sample_distribution}).
For arteries and veins, only pixels within the ROI and excluding uncertain vessels and crossings were considered, aligning with common practices in the literature~\cite{Girard_Joint_AIM_2019,Hemelings_AV_CMIG_2019,galdran2022sota,karlsson2022artery}.
This ensures a fair comparison with other works that typically disregard crossings during evaluation.
In each case, the positive class is the structure of interest, and the negative class is everything else within the ROI.

A/V and BV/BG classification performances were evaluated using one-versus-one evaluation metrics (sensitivity, specificity, and accuracy), considering arteries and BV as the positive class, respectively.
Only vessel pixels excluding crossings and uncertain vessels were involved in the calculation.
While most prior works~\cite{Girard_Joint_AIM_2019,Hemelings_AV_CMIG_2019,galdran2022sota,karlsson2022artery} only consider the intersection between predicted and GT vessels, this approach can yield misleading performance measures, especially for poor segmentations, and hinders standardized comparisons.
Therefore, following recent works~\cite{Morano_AIIM_2021,Chen_MIA_2021}, we also report classification performance for all vessel pixels in the GT, including those that were not detected as such by the models.
In our state-of-the-art comparison, we report the classification performance for both scenarios, explicitly specifying the evaluation criteria used.

Additionally, we assessed the topological connectivity of A/V segmentation maps using infeasible (INF) and correct (COR) path percentages, as introduced in \citet{araujo2019deep}\footnote{\url{https://github.com/rjtaraujo/dvae-refiner} (accessed on 2023-12-20)} and adopted by \citet{Chen_MIA_2021}.
These metrics involve randomly sampling paths from both GT and generated masks, classifying them as infeasible if they are absent in the generated mask and correct if they differ by less than 10\% from the GT.
Higher COR and lower INF values indicate more topologically accurate segmentations.

Beyond these quantitative evaluations, a qualitative evaluation was performed by visual inspection of the different segmentation maps.
In particular, we focus on manifest classification errors and vessel continuity.

\begin{table*}[t]
    \newcommand{\cmidr}{\cmidrule{2-8}}
    \centering
    \caption{Ablation study in RITE.
    W-Net: 2 stacked U-Nets without recursive refinement ($K=1$), similar to~\citet{galdran2022sota}.
    RRU-Net: recursive refinement using a single U-Net module, similar to~\citet{Mosinska_CVPR_2018}.
    RRWNetAll: proposed approach with recursive refinement ($K=6$) for arteries, veins, and BV.
    RRWNet: proposed approach with recursive refinement of arteries and veins only ($K=6$).
    A one-tailed Wilcoxon signed-rank test was performed to compare the results of the proposed RRWNet with the results of the best or second best model for each evaluation metric.
    *: $p<0.05$.
    Best results are highlighted in \textbf{bold}, and second best results are \underline{underlined}.
    }
    \label{tab:ablation}
    \resizebox{\linewidth}{!}{%
    \begin{tabular}{@{\extracolsep{4pt}}lllccccc}
    \toprule

    \textbf{Evaluation}
    & \textbf{Structure}
    & \textbf{Metric}
    & \multicolumn{5}{c}{\textbf{Models}}
    \\

    \cmidrule{4-8}

    & &
    & \multicolumn{1}{c}{U-Net}
    & \multicolumn{1}{c}{W-Net}
    & \multicolumn{1}{c}{RRU-Net}
    & \multicolumn{1}{c}{RRWNetAll}
    & \multicolumn{1}{c}{\textit{RRWNet}}
    \\

    \midrule

    \multirow{6}{*}{Segmentation}
    & \multirow{2}{*}{Artery}
    & AUROC
    & $97.13\pm0.18$
    & $97.30\pm0.20$
    & $97.37\pm0.51$
        & \underline{$97.73\pm0.20$}
    & $\mathbf{97.89\pm0.28}$*
    \\

    &
    & AUPR
    & $81.18\pm0.44$
    & $83.50\pm0.91$
    & $82.72\pm1.88$
        & \underline{$84.34\pm0.22$}
    & $\mathbf{86.60\pm0.35}$*
    \\

    \cmidr

    & \multirow{2}{*}{Vein}
    & AUROC
    & $98.00\pm0.13$
    & $97.89\pm0.12$
    & $97.99\pm0.21$
        & \underline{$98.14\pm0.10$}
    & $\mathbf{98.33\pm0.13}$*
    \\

    &
    & AUPR
    & $87.03\pm0.22$
    & $87.94\pm0.48$
    & $87.74\pm1.12$
        & \underline{$88.32\pm0.47$}
    & $\mathbf{90.14\pm0.30}$*
    \\

    \cmidr

    & \multirow{2}{*}{BV}
    & AUROC
    & $98.24\pm0.03$
    & $98.23\pm0.06$
    & \underline{$98.32\pm0.03$}
    & $98.11\pm0.11$
    & $\mathbf{98.46\pm0.05}$*
    \\

    &
    & AUPR
    & $92.61\pm0.09$
    & $92.66\pm0.05$
    & \underline{$92.86\pm0.12$}
    & $92.08\pm0.18$
    & $\mathbf{93.18\pm0.05}$*
    \\

    \midrule

    \multirow{6}{*}{Classification}
    & \multirow{3}{*}{Artery/Vein}
    & Sens.
    & $86.54\pm1.85$
    & $90.92\pm0.34$
    & $89.96\pm2.08$
    & \underline{$92.97\pm0.91$}
    & $\mathbf{94.00\pm0.50}$*
    \\

    &
    & Spec.
    & $91.27\pm0.66$
    & $92.31\pm1.49$
    & $92.15\pm1.35$
    & \underline{$93.95\pm0.72$}
    & $\mathbf{95.16\pm0.27}$*
    \\

    &
    & Acc.
    & $89.14\pm0.67$
    & $91.68\pm0.75$
    & $91.16\pm1.09$
    & \underline{$93.51\pm0.59$}
    & $\mathbf{94.63\pm0.24}$*
    \\

    \cmidr

    & \multirow{3}{*}{BV/BG}
    & Sens.
    & $81.63\pm1.37$
    & \underline{$81.85\pm3.22$}
    & $\mathbf{82.56\pm2.48}$
    & $80.30\pm3.69$
    & $81.51\pm2.25$
    \\

    &
    & Spec.
    & \underline{$98.27\pm0.20$}
    & $98.18\pm0.46$
    & $98.15\pm0.40$
    & $98.23\pm0.52$
    & $\mathbf{98.41\pm0.32}$
    \\

    &
    & Acc.
    & $96.17\pm0.01$
    & $96.12\pm0.03$
    & \underline{$96.18\pm0.05$}
    & $95.97\pm0.08$
    & $\mathbf{96.28\pm0.02}$*
    \\

    \bottomrule
    \end{tabular}%
    }
\end{table*}

\subsection{Training and evaluation details}

We employed 4-fold cross-validation on the RITE and HRF training sets, dividing each fold into 80\% for training and 20\% for validation.
The Adam optimizer~\cite{Kingma_Adam_ICLR_2015} with a constant learning rate of $\alpha = 1 \times 10^{-4}$ and decay rates of $\beta_1 = 0.9$ and $\beta_2 = 0.999$ was used for training.
Early stopping was applied after 200 epochs with no decrease in validation loss.
The batch size is set to 1.
For the state-of-the-art comparison, models with the lowest validation error among the different folds were chosen.

We maintained the original splits of RITE and HRF for training and testing.
RITE images were used at full resolution for both training and testing.
HRF images were resized to 1024 pixels wide for training and testing, but predicted segmentation maps were then upsampled to the original resolution for the evaluation, following \citet{galdran2022sota}.
Similarly, for LES-AV, we predicted at full resolution by feeding the model trained on RITE with LES-AV images resized to 576 pixels wide and then upsampling the predictions.

All images underwent offline preprocessing including global contrast enhancement and local intensity normalization, following \citet{Morano_AIIM_2021}.
Online data augmentation with color/intensity variations, affine transformations, flipping, and random cutout was applied during training.

For INF and COR calculations, 1000 paths were used for RITE and 100 for HRF and LES-AV datasets, balancing computational cost with metric reliability.

The methodology was implemented in Python 3 with PyTorch.
The code, model weights, and test set predictions will be available on GitHub: \url{https://github.com/j-morano/rrwnet}.
The experiments were run on a server with dual AMD EPYC 7443 24-Core CPUs (1024GB of RAM) and one NVIDIA RTX A6000 GPU.
Training RRWNet takes approximately 3 hours on this setup, while image segmentation takes under 0.1 seconds on the GPU and 6-8 seconds on the CPU.

\section{Results and Discussion}

\subsection{Hyperparameter search}

Table~\ref{tab:search_k} shows the AUROC and AUPR values for A/V/BV segmentation, as well as the mean sensitivity, specificity, and accuracy values for A/V classification and BV/BG classification in RITE (validation) for the RRWNet model with different $K$.
These results show that the proposed method exhibits robustness to the choice of this hyperparameter, achieving comparable performance across all $K$ values.
However, with $K=6$, the model achieved slightly better results in 3 out of 12 metrics and was the second-best in 5 others.
Notably, the mean of AUROC, AUPR, and accuracy for segmentation and classification for $K=6$ ($94.16\pm4.40$) is slightly higher than for other $K$ values
(2: $93.79\pm4.64$, 3: $94.12\pm4.46$, 8: $94.12\pm4.42$, and 11: $94.13\pm4.40$).
Based on these findings, we selected $K=6$ for the remaining experiments.

\begin{figure*}
    \centering
    \begin{tabularx}{\textwidth}{ZZZ}
        GT & Base segmentation ($k=0$) & $k=1$ \\
        \includegraphics[width=0.30\textwidth]{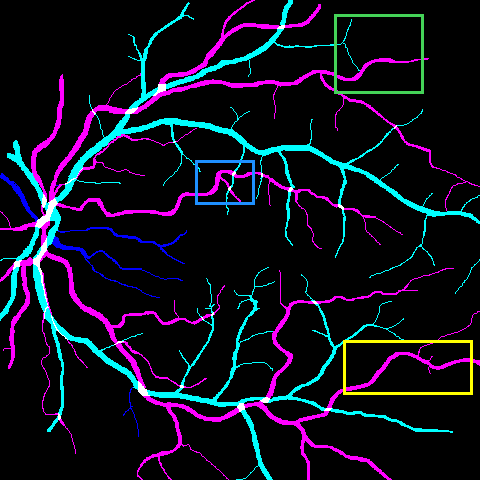}
        & \includegraphics[width=0.30\textwidth]{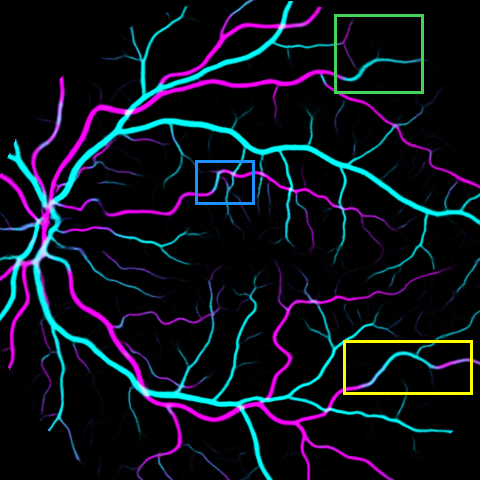}
        & \includegraphics[width=0.30\textwidth]{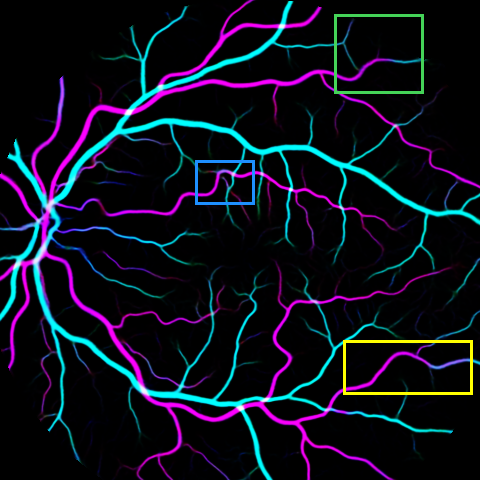} \\

        \includegraphics[height=0.06\textwidth]{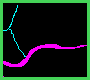}
        \includegraphics[height=0.06\textwidth]{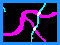}
        \includegraphics[height=0.06\textwidth]{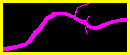}
        & \includegraphics[height=0.061\textwidth]{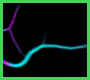}
        \includegraphics[height=0.06\textwidth]{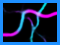}
        \includegraphics[height=0.06\textwidth]{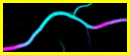}
        & \includegraphics[height=0.06\textwidth]{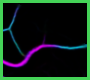}
        \includegraphics[height=0.06\textwidth]{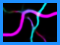}
        \includegraphics[height=0.06\textwidth]{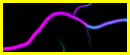} \\

        $k=2$ & $k=3$ & $k=4$ \\
        \includegraphics[width=0.30\textwidth]{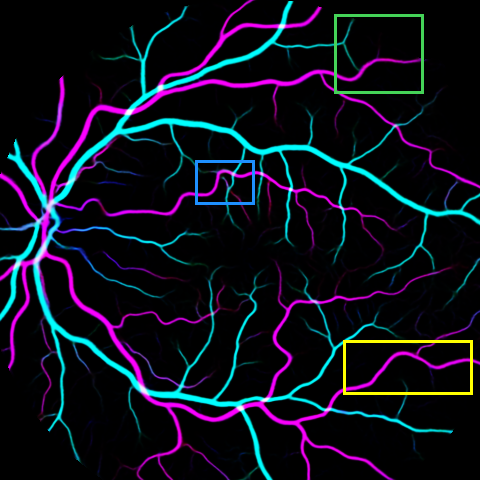}
        & \includegraphics[width=0.30\textwidth]{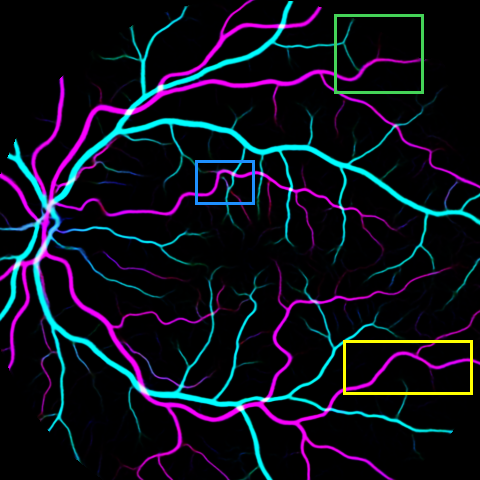}
        & \includegraphics[width=0.30\textwidth]{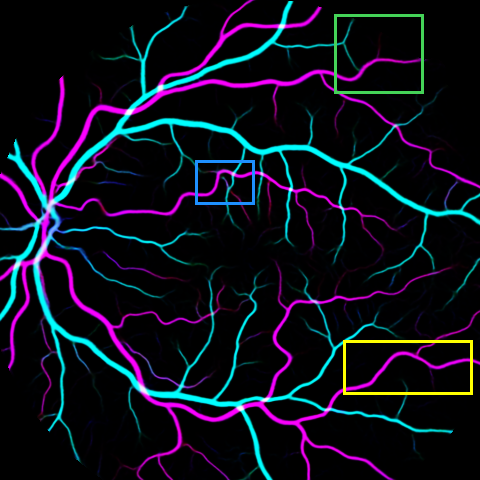} \\
        \includegraphics[height=0.06\textwidth]{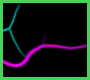}
        \includegraphics[height=0.06\textwidth]{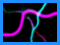}
        \includegraphics[height=0.06\textwidth]{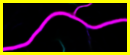}
        & \includegraphics[height=0.06\textwidth]{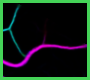}
        \includegraphics[height=0.06\textwidth]{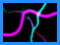}
        \includegraphics[height=0.06\textwidth]{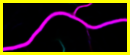}
        & \includegraphics[height=0.06\textwidth]{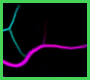}
        \includegraphics[height=0.06\textwidth]{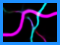}
        \includegraphics[height=0.06\textwidth]{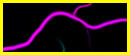} \\

        $k=5$ & $k=6$ & \\
        \includegraphics[width=0.30\textwidth]{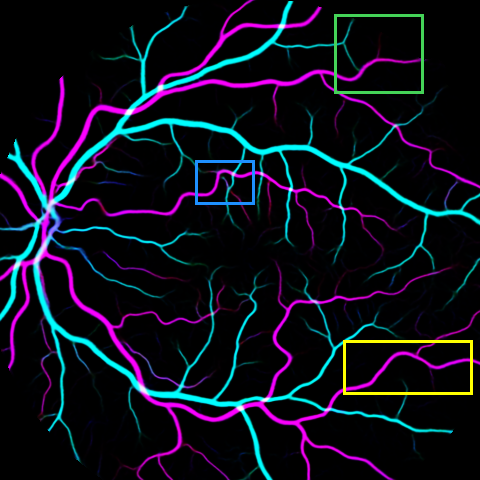}
        & \includegraphics[width=0.30\textwidth]{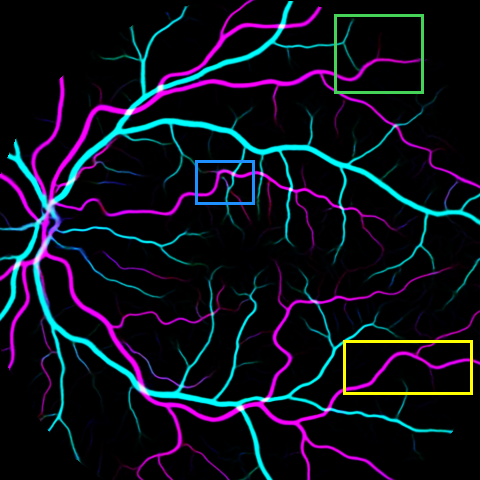}
        &
        \\

        \includegraphics[height=0.06\textwidth]{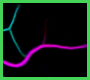}
        \includegraphics[height=0.06\textwidth]{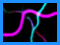}
        \includegraphics[height=0.06\textwidth]{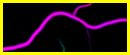}
        & \includegraphics[height=0.06\textwidth]{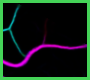}
        \includegraphics[height=0.06\textwidth]{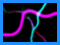}
        \includegraphics[height=0.06\textwidth]{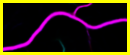}
        &
        \\

    \end{tabularx}
    \caption{
        Effect of the refinement module of RRWNet on A/V classification.
        The iterative refinement approach progressively improves A/V classification.
        For this particular example, the initial accuracy of 86.86\% ($k=0$) is improved to an accuracy of 88.89\% ($k=6$).
        [RITE, image \texttt{05}]
    }
    \label{fig:k_effect}
\end{figure*}

Fig.~\ref{fig:k_effect} displays the segmentation maps generated by RRWNet ($K=6$) at each iteration $k$.
As evident in the figure, the RR module progressively improves A/V classification over iterations.
Notably, in the base segmentation map ($k=0$), several misclassified vessels are visible.
However, these errors are progressively reduced in subsequent iterations, resulting in a final segmentation map ($k=6$) with significantly fewer manifest classification errors and improved delineation of both arteries and veins.

\begin{figure*}[t]
    \captionsetup[subfigure]{labelformat=empty}
    \centering
    \begin{tabularx}{\textwidth}{ZZZ}
    GT & U-Net & W-Net \\

    \includegraphics[width=0.32\textwidth]
        {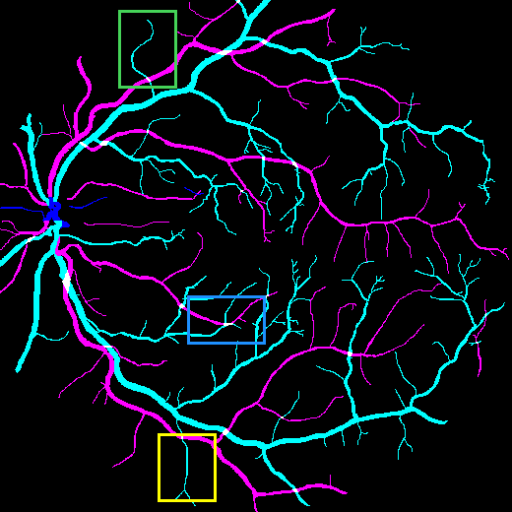}
    & \includegraphics[width=0.32\textwidth]
        {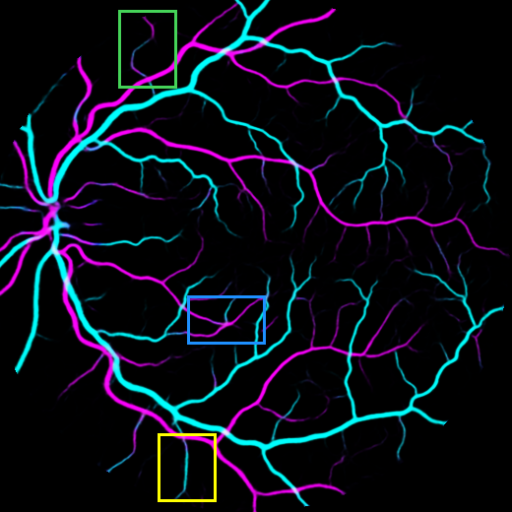}
    & \includegraphics[width=0.32\textwidth]
        {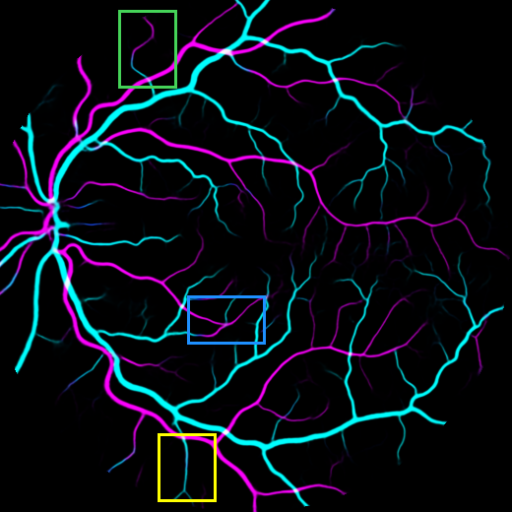} \\

    \includegraphics[height=0.06\textwidth]
        {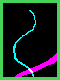}
    \includegraphics[height=0.06\textwidth]
        {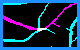}
    \includegraphics[height=0.06\textwidth]
        {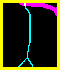}
    & \includegraphics[height=0.06\textwidth]
        {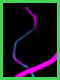}
    \includegraphics[height=0.06\textwidth]
        {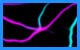}
    \includegraphics[height=0.06\textwidth]
        {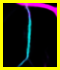}
    & \includegraphics[height=0.06\textwidth]
        {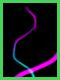}
    \includegraphics[height=0.06\textwidth]
        {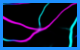}
    \includegraphics[height=0.06\textwidth]
        {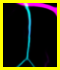} \\~\\

    RR-UNet & RRWNetAll & RRWNet \\

    \includegraphics[width=0.32\textwidth]
        {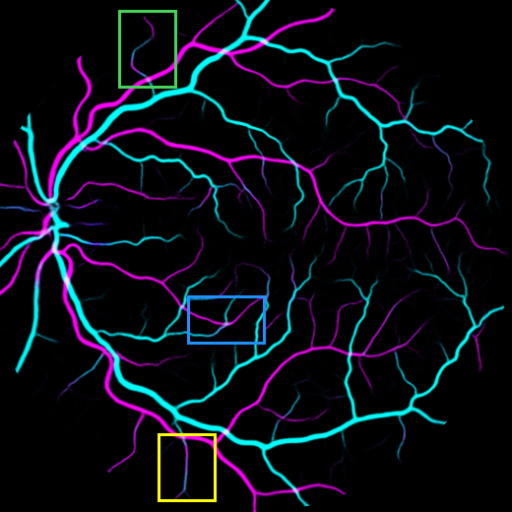}
    & \includegraphics[width=0.32\textwidth]
        {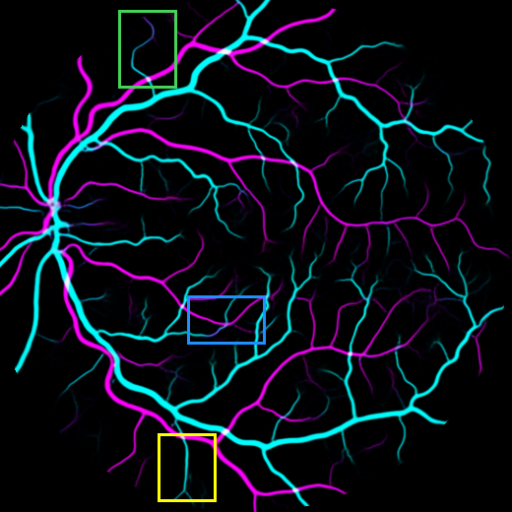}
    & \includegraphics[width=0.32\textwidth]
        {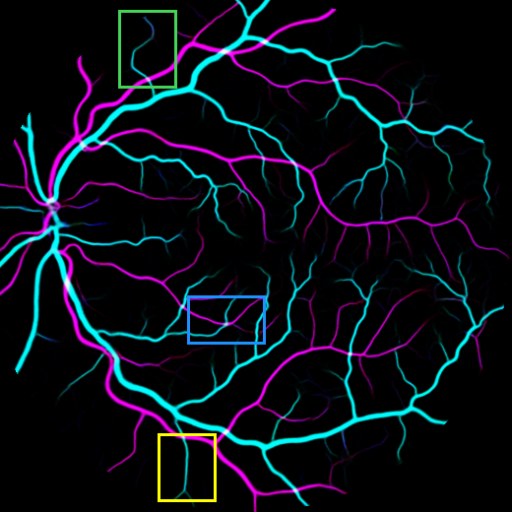} \\

    \includegraphics[height=0.06\textwidth]
        {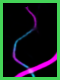}
    \includegraphics[height=0.06\textwidth]
        {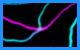}
    \includegraphics[height=0.06\textwidth]
        {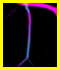}
    & \includegraphics[height=0.06\textwidth]
        {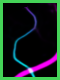}
    \includegraphics[height=0.06\textwidth]
        {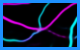}
    \includegraphics[height=0.06\textwidth]
        {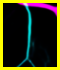}
    & \includegraphics[height=0.06\textwidth]
        {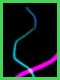}
    \includegraphics[height=0.06\textwidth]
        {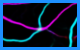}
    \includegraphics[height=0.06\textwidth]
        {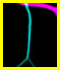} \\

    \end{tabularx} \\
    \caption{
        Examples of segmentation maps obtained by the different models in RITE.
        Some differences between the segmentation maps of the different models are highlighted in colored boxes.
        [RITE, image \texttt{09}]
    }
    \label{fig:examples_RITE_1}
\end{figure*}

\begin{table*}[tb]
    \centering
    \caption{
        Comparison with the state of the art in the tasks of A/V classification and vessel segmentation.
        Here, only detected vessels are considered.
        All values are percentages.
        Highest values among the automatic methods for each metric and dataset are highlighted in \textbf{bold}.
        All (*) or BV segmentation results (\textdagger) calculated by us in the absence of reported values but available code/predictions.
        $\ddagger$ Cross-dataset evaluation (trained on RITE).
        $\mathsection$ 2-fold cross-validation.
        $\mathparagraph$ Centerline-based evaluation.
    }
    \label{tab:SOTA_clasificacion}
    \resizebox{\textwidth}{!}{%
    \begin{tabular}{@{\extracolsep{4pt}}lllllllll}

    \toprule

    \textbf{Dataset}
    & \textbf{Method}
    & \multicolumn{3}{c}{A/V classification}
    & \multicolumn{4}{c}{BV segmentation (BV/BG classification)}
    \\

    \cmidrule{3-5} \cmidrule{6-9}

    &
    & Sens.
    & Spec.
    & Acc.
    & Sens.
    & Spec.
    & Acc.
    & AUROC
    \\

    \midrule

    \multirow{1}{*}{RITE}
    & \citet{Girard_Joint_AIM_2019}
    & 86.3
    & 86.6
    & 86.5
    & 78.4
    & 98.1
    & 95.7
    & 97.2
    \\

    & \citet{Galdran_Uncertainty_ISBI_2019}
    & 89
    & 90
    & 89
    & \textbf{94}
    & 93
    & 93
    & 95
    \\

    & \citet{Ma_MultiTask_MICCAI_2019}
    & 93.4
    & 95.5
    & 94.5
    & 79.16
    & 98.11
    & 95.70
    & 98.10
    \\

    & \citet{Hemelings_AV_CMIG_2019}*
    & 95.13
    & 92.78
    & 93.81
    & 77.61
    & \textbf{98.74}
    & 96.08
    & 88.17
    \\

    & \citet{Kang_AV_CMPB_2020}
    & 88.63
    & 92.72
    & 90.81
    & -
    & -
    & -
    & -
    \\

    & \citet{Morano_AIIM_2021}
    & 87.47
    & 90.89
    & 89.24
    & 79.12
    & 98.65
    & 96.16
    & 98.33
    \\

    & \citet{galdran2022sota}*
    & 88.86
    & 96.04
    & 92.76
    & 83.05
    & 98.19
    & \textbf{96.29}
    & 98.47
    \\

    & \citet{hatamizadeh2022ravir}
    & 93.10
    & 94.31
    & 95.13
    & -
    & -
    & -
    & - \\

    & \citet{karlsson2022artery}
    & 95.1
    & 96.0
    & 95.6
    & 82.2
    & 97.6
    & 95.6
    & 98.1
    \\

    & \citet{Chen_MIA_2021}\textsuperscript{\textdagger}
    & 95.38
    & 97.20
    & 96.34
    & 81.51
    & 97.81
    & 95.75
    & 96.29
    \\

    & \citet{hu2024semi}
    & 93.37
    & 95.37
    & 94.42
    & 79.08
    & 98.15
    & 95.69
    & 98.07
    \\

    & \textcolor{black!60}{Second expert~\cite{Qureshi_ISCBMS_2013}}
    & \textcolor{black!60}{95.80}
    & \textcolor{black!60}{96.82}
    & \textcolor{black!60}{96.37}
    & \textcolor{black!60}{80.38}
    & \textcolor{black!60}{96.83}
    & \textcolor{black!60}{94.76}
    & \textcolor{black!60}{-}
    \\

    & \textit{RRWNet (ours)}
    & \textbf{95.73}
    & \textbf{97.38}
    & \textbf{96.66}
    & 80.16
    & 98.61
    & \textbf{96.29}
    & \textbf{98.50}
    \\


    \midrule

    \multirow{1}{*}{LES-AV}
    & \citet{Galdran_Uncertainty_ISBI_2019}\textsuperscript{$\ddagger$}
    & 88
    & 85
    & 86
    & -
    & -
    & -
    & -
    \\

    & \citet{Kang_AV_CMPB_2020}\textsuperscript{$\mathsection$}
    & 94.26
    & 90.90
    & 92.19
    & -
    & -
    & -
    & -
    \\

    & \citet{galdran2022sota}*\textsuperscript{$\ddagger$}
    & 86.86
    & 93.56
    & 90.47
    & 76.40
    & \textbf{97.73}
    & \textbf{95.69}
    & 96.27
    \\

    & \textit{RRWNet (ours)}\textsuperscript{$\ddagger$}
    & \textbf{94.30}
    & \textbf{95.25}
    & \textbf{94.81}
    & \textbf{86.41}
    & 96.59
    & 95.61
    & \textbf{97.72}
    \\

    \midrule

    \multirow{1}{*}{HRF}
    & \citet{Galdran_Uncertainty_ISBI_2019}
    & 85
    & 91
    & 91
    & -
    & -
    & -
    & -
    \\

    & \citet{Hemelings_AV_CMIG_2019}*
    & -
    & -
    & 96.98\textsuperscript{$\mathparagraph$}
    & 80.74
    & -
    & -
    & -
    \\

    & \citet{Chen_MIA_2021}*
    & 97.06
    & 97.29
    & 97.19
    & 78.14
    & 98.29
    & 96.59
    & 94.66
    \\

    & \citet{galdran2022sota}*
    & \textbf{98.10}
    & 93.17
    & 95.35
    & 81.19
    & 98.12
    & \textbf{96.70}
    & 98.55
    \\

    & \citet{karlsson2022artery}*
    & 97.07
    & 96.53
    & 96.77
    & \textbf{86.17}
    & 97.09
    & 96.17
    & 98.42
    \\

    & \citet{hu2024semi}
    & 93.37
    & 95.37
    & 94.42
    & 69.01
    & \textbf{99.02}
    & 96.25
    & 98.15
    \\

    & \textcolor{black!60}{Second expert~\cite{Hemelings_AV_CMIG_2019}}
    & \textcolor{black!60}{97.46}
    & \textcolor{black!60}{97.05}
    & \textcolor{black!60}{97.23}
    & \textcolor{black!60}{93.85}
    & \textcolor{black!60}{98.91}
    & \textcolor{black!60}{98.48}
    & \textcolor{black!60}{-}
    \\

    & \textit{RRWNet (ours)}
    & 97.98
    & \textbf{97.72}
    & \textbf{97.83}
    & 82.78
    & 97.87
    & 96.60
    & \textbf{98.57}
    \\

    \bottomrule

    \end{tabular}%
    }
\end{table*}

\subsection{Ablation study}

Table~\ref{tab:ablation} shows the mean area under the ROC curve (AUROC) and area under the PR curve (AUPR) for A/V/BV segmentation, as well as the mean sensitivity, specificity, and accuracy values for A/V classification and BV/BG classification in RITE for different variants of the proposed RRWNet model.

All evaluated methods achieved superior segmentation performance compared to the U-Net baseline across all evaluation metrics, except W-Net and RRU-Net for vein AUROC (which showed marginal reductions of 0.11 pp and 0.01 pp, respectively) and RRU-Net for BV AUROC (0.01 pp reduction).
Interestingly, RRWNetAll, which recursively refines \textit{all} segmentation maps (A/V/BV), led to improved A/V segmentation but resulted in decreased performance for BV segmentation (with reductions of 0.13 pp and 0.53 pp in AUROC and AUPR, respectively).
Conversely, the proposed RRWNet, which focuses solely on refining A/V segmentation maps, yielded significant improvements in all segmentation tasks compared to the U-Net baseline and the other methods.
RRWNet combines the high BV segmentation performance of U-Net with the increased A/V segmentation accuracy provided by the refinement module.
These improvements were particularly notable for arteries and veins, with AUPR values exceeding those of U-Net by 5.64 pp and 3.11 pp, respectively. Similar trends were observed for AUROC values, which were 0.80 pp and 0.36 pp higher, respectively.
The improvement in terms of AUPR is particularly relevant due to its increased sensitivity compared to AUROC in scenarios with imbalanced classes, as is the case with arteries and veins in this study.

Similar to segmentation, all methods surpassed the U-Net baseline in all A/V classification metrics, with RRWNet demonstrating statistically significant superiority over the second-best method in all cases.
Notably, RRWNet outperformed U-Net by significant margins in sensitivity (+7.41 pp), specificity (+3.89 pp), and accuracy (+5.48 pp).
Similar to RRWNet, RRWNetAll achieved improved A/V classification performance compared to W-Net and RRU-Net, exhibiting an accuracy increase of 1.83 pp and 2.35 pp, respectively. Smaller improvements in A/V accuracy were observed for both W-Net (+1.54 pp) and RRU-Net (+1.02 pp) compared to the U-Net baseline.

Unlike the previous categories (A/V/BV segmentation and A/V classification), methods employing refinement strategies did not consistently outperform the U-Net baseline in BV/BG classification, with slight performance reductions observed in 7 out of 12 metrics. This decrease was particularly evident for RRWNetAll, which yielded a 0.2 pp lower accuracy compared to the U-Net baseline, consistently with its BV segmentation peformance.
Despite the observed decrease in certain metrics for some refinement methods, RRWNet remained the best performing method in terms of both sensitivity and accuracy for BV/BG classification, with significant improvements over other methods in the latter metric.
The observed discrepancy in performance between BV/BG classification and BV segmentation, where all metrics exhibited statistically significant differences, can be partially attributed to the threshold employed for binarizing the model outputs.
In BV segmentation, the evaluation metrics (AUROC and AUPR) are inherently threshold-independent, circumventing this potential issue.
However, classification metrics were computed using the binary segmentation maps obtained after applying a threshold of 0.5 to the predicted probability maps, following \citet{Morano_AIIM_2021} and \citet{karlsson2022artery}.
While this threshold optimizes accuracy on the training set, it may not generalize optimally to the test set, potentially explaining the observed performance difference.

Overall, the proposed RRWNet framework achieved superior performance compared to other ablation methods in 11 out of 12 evaluated metrics, with significant improvements observed in 10 of them. These results highlight the effectiveness of the proposed architectural design, and in particular the RR module, in improving the classification and segmentation of arteries and veins.

Figure~\ref{fig:examples_RITE_1} shows the segmentation maps generated by different models within the ablation study in RITE.
These qualitative observations corroborate the quantitative results presented above.
The proposed RRWNet, incorporating the RR module, demonstrates its ability to solve manifest classification errors. This results in segmentation maps that are more topologically accurate and exhibit greater fidelity to the GT. Notably, the model achieves this without the need for additional topology constraints or post-processing techniques, showcasing its inherent capability in addressing these issues.

\subsection{State of the art comparison}

\begin{table*}[t]
    \centering
    \caption{
        Comparison with the state of the art in artery and vein classification and segmentation.
        In this case, all ``vessel'' pixels from the GT except crossings and unknown pixels are considered for the evaluation.
        All values are in percentages.
        The results obtained by applying the proposed RR module as a post-processing step to the segmentation maps generated by the other methods are shown in parentheses.
        When this value is higher than the value obtained by the method itself, it is highlighted in \textcolor{darkgreen}{green}; when it is lower, in \textcolor{darkred}{red}.
        The best values among the automatic end-to-end methods (i.e., excluding the post-processing step) for each metric and dataset are highlighted in \textbf{bold}.
        The best overall (including both with and without the post-processing step) are \underline{underlined}.
    }
    \label{tab:av_segmentation}
    \resizebox{\textwidth}{!}{%
    \begin{tabular}{@{\extracolsep{4pt}}lllllllllll}

    \toprule

    \textbf{Dataset}
    & \textbf{Method}
    & \multicolumn{1}{c}{A/V}
    & \multicolumn{4}{c}{Artery}
    & \multicolumn{4}{c}{Vein}
    \\

    \cmidrule{3-3} \cmidrule{4-7} \cmidrule{8-11} 

    &
    & Acc.
    & AUPR
    & AUROC
    & COR
    & INF $\downarrow$
    & AUPR
    & AUROC
    & COR
    & INF $\downarrow$
    \\

    \midrule

    RITE
    & \citet{Morano_AIIM_2021}
    & 89.26 {\textcolor{darkgreen}{\footnotesize(94.37)}}
    & 81.49 {\textcolor{darkgreen}{\footnotesize(86.46)}}
    & 97.37 {\textcolor{darkgreen}{\footnotesize(97.95)}}
    & 13.71 {\textcolor{darkgreen}{\footnotesize(28.79)}}
    & 86.02 {\textcolor{darkgreen}{\footnotesize(70.84)}}
    & 87.26 {\textcolor{darkred}{\footnotesize(86.97)}}
    & 98.12 {\textcolor{darkgreen}{\footnotesize(98.16)}}
    & 27.54 {\textcolor{darkgreen}{\footnotesize(40.95)}}
    & 72.23 {\textcolor{darkgreen}{\footnotesize(58.74)}}
    \\

    & \citet{galdran2022sota}
    & 90.59 {\textcolor{darkgreen}{\footnotesize(94.80)}}
    & 83.26 {\textcolor{darkgreen}{\footnotesize(\underline{87.21})}}
    & 97.31 {\textcolor{darkgreen}{\footnotesize(98.02)}}
    & 9.93  {\textcolor{darkgreen}{\footnotesize(27.15)}}
    & 89.85 {\textcolor{darkgreen}{\footnotesize(72.48)}}
    & 87.71 {\textcolor{darkred}{\footnotesize(84.81)}}
    & 98.21 {\textcolor{darkgreen}{\footnotesize(98.27)}}
    & 16.84 {\textcolor{darkgreen}{\footnotesize(39.14)}}
    & 82.84 {\textcolor{darkgreen}{\footnotesize(60.47)}}
    \\

    & \citet{karlsson2022artery}
    & 94.67 {\textcolor{darkgreen}{\footnotesize(94.70)}}
    & 86.39 {\textcolor{darkgreen}{\footnotesize(86.81)}}
    & 97.79 {\textcolor{darkgreen}{\footnotesize(97.62)}}
    & 14.42 {\textcolor{darkgreen}{\footnotesize(\underline{33.19})}}
    & 85.30 {\textcolor{darkgreen}{\footnotesize(\underline{66.42})}}
    & 89.47 {\textcolor{darkred}{\footnotesize(84.47)}}
    & 98.27 {\textcolor{darkred}{\footnotesize(97.76)}}
    & 22.39 {\textcolor{darkgreen}{\footnotesize(42.91)}}
    & 77.35 {\textcolor{darkgreen}{\footnotesize(56.75)}}
    \\

    & \citet{Chen_MIA_2021}
    & 90.91 {\textcolor{darkgreen}{\footnotesize(93.93)}}
    & 80.94 {\textcolor{darkgreen}{\footnotesize(84.03)}}
    & 94.81 {\textcolor{darkgreen}{\footnotesize(96.49)}}
    & 19.04 {\textcolor{darkgreen}{\footnotesize(29.84)}}
    & 80.56 {\textcolor{darkgreen}{\footnotesize(69.80)}}
    & 85.75 {\textcolor{darkred}{\footnotesize(85.50)}}
    & 95.18 {\textcolor{darkgreen}{\footnotesize(97.64)}}
    & 25.16 {\textcolor{darkgreen}{\footnotesize(\underline{46.78})}}
    & 74.67 {\textcolor{darkgreen}{\footnotesize(\underline{52.68})}}
    \\

    & \textit{RRWNet (ours)}
    & \textbf{\underline{94.95}}
    & \textbf{86.93}
    & \textbf{\underline{98.22}}
    & \textbf{31.62}
    & \textbf{68.03}
    & \textbf{\underline{90.43}}
    & \textbf{\underline{98.31}}
    & \textbf{38.23}
    & \textbf{61.36}
    \\

    \midrule

    LES-AV
    & \citet{Morano_AIIM_2021}
    & 83.62 {\textcolor{darkgreen}{\footnotesize(88.44)}}
    & 72.45 {\textcolor{darkgreen}{\footnotesize(78.02)}}
    & 96.64 {\textcolor{darkgreen}{\footnotesize(97.73)}}
    & 11.73 {\textcolor{darkgreen}{\footnotesize(26.00)}}
    & 87.82 {\textcolor{darkgreen}{\footnotesize(73.77)}}
    & 80.53 {\textcolor{darkgreen}{\footnotesize(80.60)}}
    & 97.48 {\textcolor{darkred}{\footnotesize(96.52)}}
    & 25.91 {\textcolor{darkgreen}{\footnotesize(32.55)}}
    & 73.64 {\textcolor{darkgreen}{\footnotesize(66.91)}}
    \\

    & \citet{galdran2022sota}
    & 85.39 {\textcolor{darkgreen}{\footnotesize(89.41)}}
    & 74.66 {\textcolor{darkgreen}{\footnotesize(79.74)}}
    & 97.08 {\textcolor{darkgreen}{\footnotesize(\underline{97.99})}}
    & 10.05 {\textcolor{darkgreen}{\footnotesize(23.68)}}
    & 89.68 {\textcolor{darkgreen}{\footnotesize(75.68)}}
    & 80.46 {\textcolor{darkgreen}{\footnotesize(83.47)}}
    & 97.20 {\textcolor{darkred}{\footnotesize(97.10)}}
    & 22.50 {\textcolor{darkgreen}{\footnotesize(35.32)}}
    & 76.86 {\textcolor{darkgreen}{\footnotesize(63.27)}}
    \\

    & \textit{RRWNet (ours)}
    & \textbf{\underline{92.61}}
    & \textbf{\underline{81.87}}
    & \textbf{97.18}
    & \textbf{\underline{47.05}}
    & \textbf{\underline{51.68}}
    & \textbf{\underline{86.50}}
    & \textbf{\underline{97.70}}
    & \textbf{\underline{49.68}}
    & \textbf{\underline{49.45}}
    \\

    \midrule

    HRF
    & \citet{Morano_AIIM_2021}
    & 94.76 {\textcolor{darkgreen}{\footnotesize(96.32)}}
    & 84.02 {\textcolor{darkgreen}{\footnotesize(84.64)}}
    & 98.86 {\textcolor{darkgreen}{\footnotesize(99.01)}}
    & 44.87 {\textcolor{darkgreen}{\footnotesize(51.80)}}
    & 54.73 {\textcolor{darkgreen}{\footnotesize(47.87)}}
    & 87.85 {\textcolor{darkred}{\footnotesize(83.83)}}
    & 98.96 {\textcolor{darkgreen}{\footnotesize(99.07)}}
    & 46.27 {\textcolor{darkred}{\footnotesize(46.13)}}
    & 53.20 {\textcolor{darkgreen}{\footnotesize(53.07)}}
    \\

    & \citet{galdran2022sota}
    & 93.94 {\textcolor{darkgreen}{\footnotesize(96.65)}}
    & 82.65 {\textcolor{darkgreen}{\footnotesize(85.26)}}
    & \textbf{98.91} {\textcolor{darkgreen}{\footnotesize(\underline{99.06})}}
    & 17.07 {\textcolor{darkgreen}{\footnotesize(55.80)}}
    & 82.33 {\textcolor{darkgreen}{\footnotesize(43.40)}}
    & 86.94 {\textcolor{darkgreen}{\footnotesize(\underline{88.64})}}
    & 98.76 {\textcolor{darkgreen}{\footnotesize(\underline{99.04})}}
    & 18.07 {\textcolor{darkgreen}{\footnotesize(52.07)}}
    & 81.47 {\textcolor{darkgreen}{\footnotesize(47.07)}}
    \\

    & \citet{karlsson2022artery}
    & 95.80 {\textcolor{darkgreen}{\footnotesize(\underline{96.91})}}
    & 83.27 {\textcolor{darkred}{\footnotesize(82.60)}}
    & 98.55 {\textcolor{darkred}{\footnotesize(98.52)}}
    & 31.80 {\textcolor{darkgreen}{\footnotesize(\underline{60.87})}}
    & 67.93 {\textcolor{darkgreen}{\footnotesize(\underline{38.40})}}
    & 86.42 {\textcolor{darkred}{\footnotesize(83.68)}}
    & 98.41 {\textcolor{darkgreen}{\footnotesize(98.50)}}
    & 23.60 {\textcolor{darkgreen}{\footnotesize(\underline{54.20})}}
    & 75.67 {\textcolor{darkgreen}{\footnotesize(\underline{45.00})}}
    \\

    & \citet{Chen_MIA_2021}
    & 92.08 {\textcolor{darkgreen}{\footnotesize(96.72)}}
    & 77.95 {\textcolor{darkgreen}{\footnotesize(81.60)}}
    & 93.75 {\textcolor{darkgreen}{\footnotesize(98.06)}}
    & 27.20 {\textcolor{darkgreen}{\footnotesize(48.87)}}
    & 72.67 {\textcolor{darkgreen}{\footnotesize(50.73)}}
    & 82.12 {\textcolor{darkgreen}{\footnotesize(82.46)}}
    & 94.58 {\textcolor{darkgreen}{\footnotesize(97.99)}}
    & 36.33 {\textcolor{darkgreen}{\footnotesize(44.47)}}
    & 63.53 {\textcolor{darkgreen}{\footnotesize(54.93)}}
    \\

    & \textit{RRWNet (ours)}
    & \textbf{95.85}
    & \textbf{\underline{84.99}}
    & \textbf{98.91}
    & \textbf{48.40}
    & \textbf{51.00}
    & \textbf{88.36}
    & \textbf{98.99}
    & \textbf{48.13}
    & \textbf{51.40}
    \\

    \bottomrule

    \end{tabular}%
    }
\end{table*}

\subsubsection{A/V classification and BV segmentation}

Table~\ref{tab:SOTA_clasificacion} presents a comparison of the performance of the proposed RRWNet model against current state-of-the-art approaches for A/V classification and BV segmentation on the RITE, LES-AV and HRF datasets.
RRWNet consistently achieved state-of-the-art performance across all datasets and most evaluation metrics considered in Table~\ref{tab:SOTA_clasificacion}.

In RITE, RRWNet achieved an A/V classification accuracy of 96.66\% and a BV segmentation AUROC of 98.50\%.
These results surpass the second-best methods, \citet{Chen_MIA_2021} and \citet{Morano_AIIM_2021}, by 0.32 pp and 0.27 pp, respectively.
Furthermore, it exceed the performance of the Second Expert~\cite{Qureshi_ISCBMS_2013} in terms of A/V and BV/BG classification by 0.29 pp and 1.53 pp, respectively, demonstrating human-level performance in both tasks.

The proposed RRWNet also achieved state-of-the-art performance in the LES-AV dataset, with an A/V classification accuracy of 94.81\% and a BV segmentation AUROC of 97.72\%.
These values represent improvements of 2.62 pp and 1.63 pp over the second best performing methods, \citet{Kang_AV_CMPB_2020} and \citet{galdran2022sota}, respectively.
Notably, RRWNet was evaluated in a cross-dataset setting (trained on RITE, tested on LES-AV),
while \citet{Kang_AV_CMPB_2020} was evaluated in a 2-fold cross-validation setting within LES-AV.
This showcases the robustness of RRWNet and its superior performance to generalize to unseen datasets.

Finally, in HRF, RRWNet achieved once again state-of-the-art performance, with an A/V classification Accuracy of 97.83\%
(+0.64 pp over \citet{Chen_MIA_2021}) and BV segmentation AUROC of 98.57\% (+0.35 pp over \citet{galdran2022sota}).

It is noteworthy that these state-of-the-art results were obtained using a straightforward implementation of RRWNet, requiring almost no hyperparameter tuning or additional post-processing steps. This further underscores the efficacy and robustness of the proposed framework.

\subsubsection{A/V segmentation and classification for all GT vessels and RR post-processing}

In addition to the standard state-of-the-art comparison presented in Table~\ref{tab:SOTA_clasificacion}, Table~\ref{tab:av_segmentation} offers a more comprehensive analysis of the performance of the proposed model compared to existing approaches.
In particular, for A/V classification, the comparison is performed in terms of accuracy (Acc.), considering all ``vessel'' pixels from the GT except crossings and unknown pixels 
For A/V segmentation, the comparison is performed using different threshold-agnostic metrics (AUPR and AUROC) and topological metrics (COR and INF).
The table also includes, in parentheses, the results obtained by applying the proposed RR module as a post-processing step to the segmentation maps generated by the other methods.
This provides insight into the potential benefit of the RR module for enhancing existing approaches.

RRWNet consistently outperformed state-of-the-art methods on all datasets and metrics considered in Table~\ref{tab:av_segmentation}.
In RITE, RRWNet achieves 94.95\% A/V accuracy, outperforming all other methods by at least 0.28 pp.
Similar improvements are observed in AUPR and AUROC for both artery and vein segmentation.
However, the most significant improvements are observed for the metrics measuring topological consistency: COR and INF.
RRWNet achieves 31.62\% COR and 68.03\% INF for arteries and 38.23\% COR and 61.36\% INF for veins.
These values represent substantial advancements, with COR being 12.58 pp higher and INF 12.53 pp lower (for INF, lower is better) for arteries and 11.69 pp higher and 10.87 pp lower for veins compared to the second-best method.
This underlines the superior ability of RRWNet to generate more topologically correct segmentation maps compared to the state of the art.
The results are similar for the HRF dataset, and even more remarkable for the LES-AV dataset, where RRWNet greatly outperforms the state-of-the-art methods in all metrics.
The differences are, again, particularly pronounced for the topological metrics, with RRWNet achieving 47.05\% COR and 51.68\% INF for arteries (+35.31 pp and -36.14 pp over the second-best method, respectively)
and 49.68\% COR and 49.45\% INF for veins (+23.77 pp and -24.46 pp over the second-best method, respectively).
This showcases the generalization capabilities of RRWNet to different an unseen datasets, and its salient ability to generate topologically correct segmentation maps.

\begin{figure*}[p]
    \captionsetup[subfigure]{labelformat=empty}
    \centering
    \begin{tabularx}{\textwidth}{ZZZ}
    GT & \textit{RRWNet (ours)} & \citet{Morano_AIIM_2021} \\

    \includegraphics[width=0.32\textwidth]
        {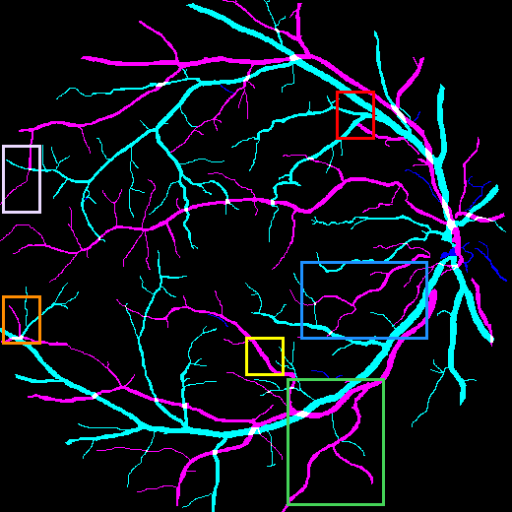}
    & \includegraphics[width=0.32\textwidth]
        {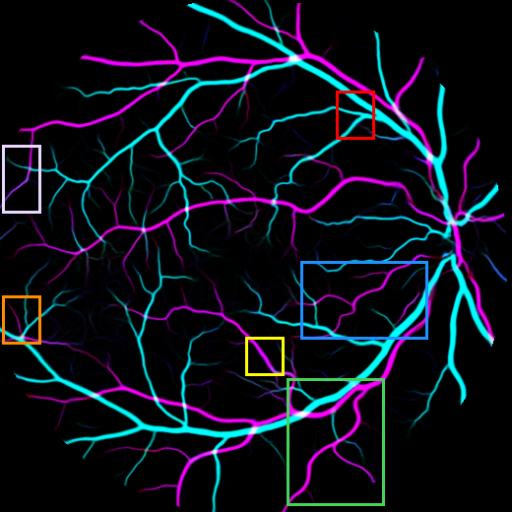}
    & \includegraphics[width=0.32\textwidth]
        {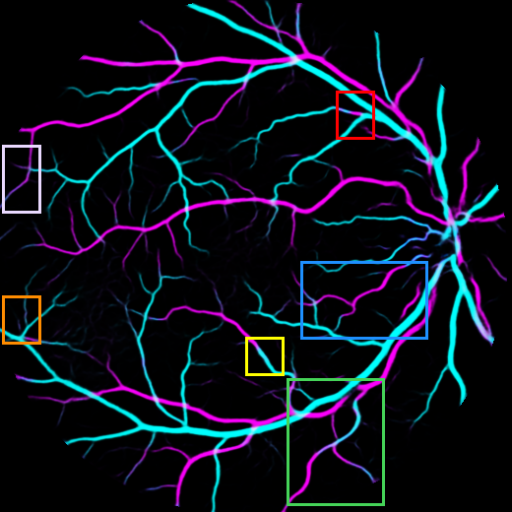}
    \\

    \includegraphics[height=0.07\textwidth]
        {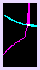}
    \includegraphics[height=0.07\textwidth]
        {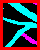}
    \includegraphics[height=0.07\textwidth]
        {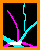}
    \includegraphics[height=0.07\textwidth]
        {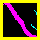}
    \includegraphics[height=0.10\textwidth]
        {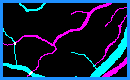}
    \includegraphics[height=0.10\textwidth]
        {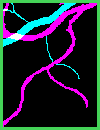}
    &
    \includegraphics[height=0.07\textwidth]
        {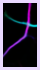}
    \includegraphics[height=0.07\textwidth]
        {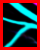}
    \includegraphics[height=0.07\textwidth]
        {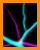}
    \includegraphics[height=0.07\textwidth]
        {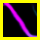}
    \includegraphics[height=0.10\textwidth]
        {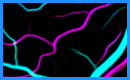}
    \includegraphics[height=0.10\textwidth]
        {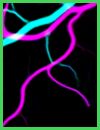}
    &
    \includegraphics[height=0.07\textwidth]
        {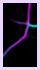}
    \includegraphics[height=0.07\textwidth]
        {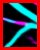}
    \includegraphics[height=0.07\textwidth]
        {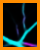}
    \includegraphics[height=0.07\textwidth]
        {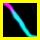}
    \includegraphics[height=0.10\textwidth]
        {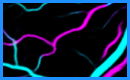}
    \includegraphics[height=0.10\textwidth]
        {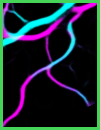}
    \\~\\

    \citet{Chen_MIA_2021} & \citet{galdran2022sota} & \citet{karlsson2022artery} \\

    \includegraphics[width=0.32\textwidth]
        {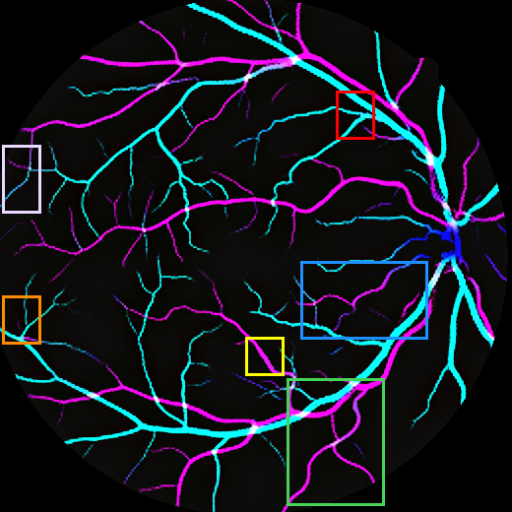}
    & \includegraphics[width=0.32\textwidth]
        {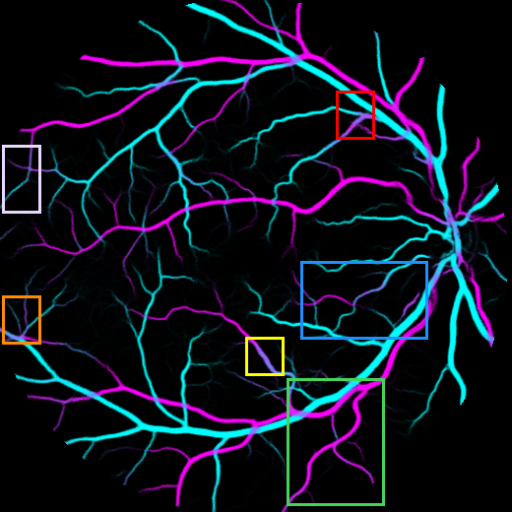}
    & \includegraphics[width=0.32\textwidth]
        {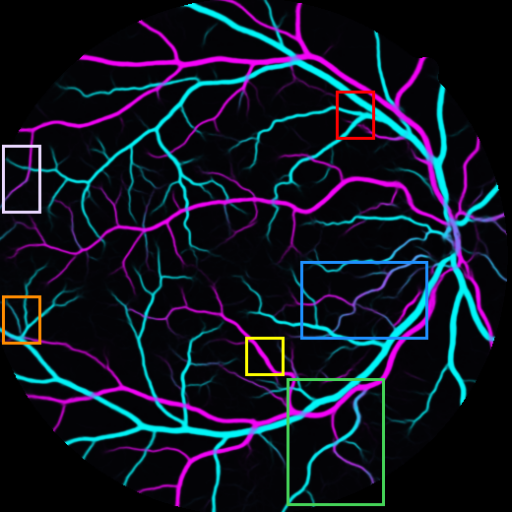}
    \\

    \includegraphics[height=0.07\textwidth]
        {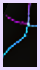}
    \includegraphics[height=0.07\textwidth]
        {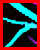}
    \includegraphics[height=0.07\textwidth]
        {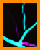}
    \includegraphics[height=0.07\textwidth]
        {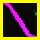}
    \includegraphics[height=0.10\textwidth]
        {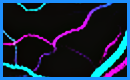}
    \includegraphics[height=0.10\textwidth]
        {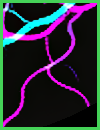}
    &
    \includegraphics[height=0.07\textwidth]
        {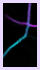}
    \includegraphics[height=0.07\textwidth]
        {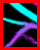}
    \includegraphics[height=0.07\textwidth]
        {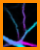}
    \includegraphics[height=0.07\textwidth]
        {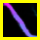}
    \includegraphics[height=0.10\textwidth]
        {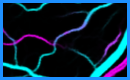}
    \includegraphics[height=0.10\textwidth]
        {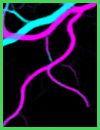}
    &
    \includegraphics[height=0.07\textwidth]
        {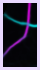}
    \includegraphics[height=0.07\textwidth]
        {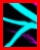}
    \includegraphics[height=0.07\textwidth]
        {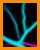}
    \includegraphics[height=0.07\textwidth]
        {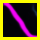}
    \includegraphics[height=0.10\textwidth]
        {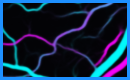}
    \includegraphics[height=0.10\textwidth]
        {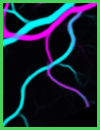} \\

    \end{tabularx} \\
    \caption{
        Examples of segmentation maps obtained by the different segmentation models in RITE dataset.
        A few notable differences between the segmentation maps of the different models are highlighted in colored boxes.
        [RITE, image \texttt{20}]
    }
    \label{fig:examples_sota_RITE_1}
\end{figure*}
\begin{figure*}[t]
    \captionsetup[subfigure]{labelformat=empty}
    \centering
    \begin{tabularx}{\textwidth}{ZZZ}
    GT & \textit{RRWNet (ours)} & \citet{Morano_AIIM_2021} \\

    \includegraphics[width=0.32\textwidth]
        {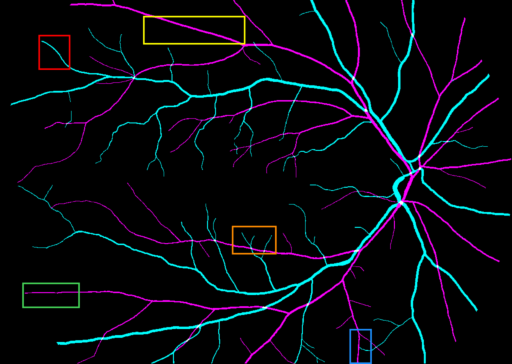}
    & \includegraphics[width=0.32\textwidth]
        {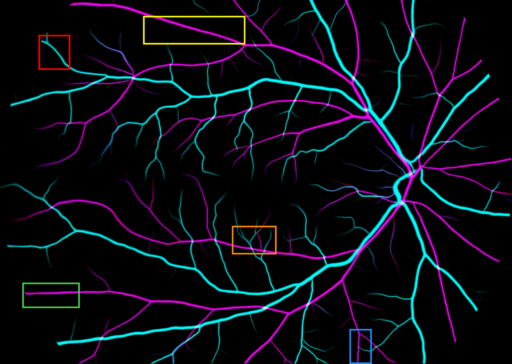}
    & \includegraphics[width=0.32\textwidth]
        {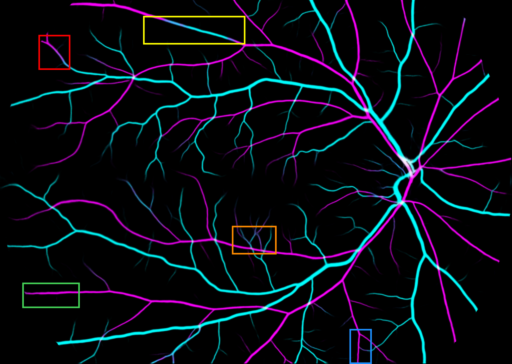}
    \\

    \includegraphics[height=0.065\textwidth]
        {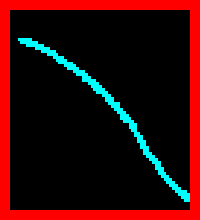}
    \includegraphics[height=0.065\textwidth]
        {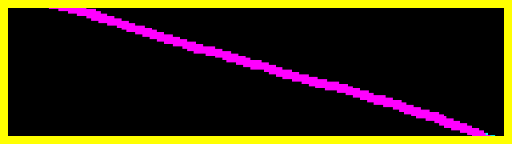}
    \includegraphics[height=0.065\textwidth]
        {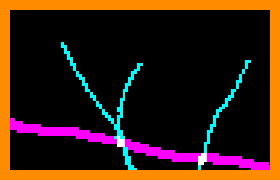}
    \includegraphics[height=0.065\textwidth]
        {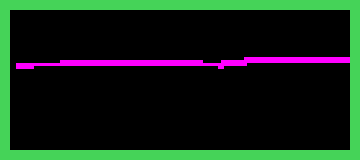}
    \includegraphics[height=0.065\textwidth]
        {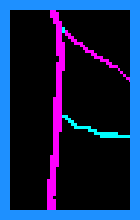}
    &
    \includegraphics[height=0.065\textwidth]
        {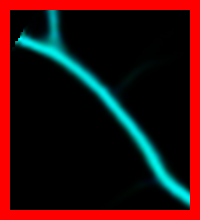}
    \includegraphics[height=0.065\textwidth]
        {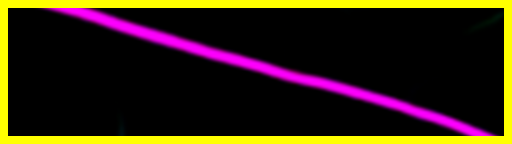}
    \includegraphics[height=0.065\textwidth]
        {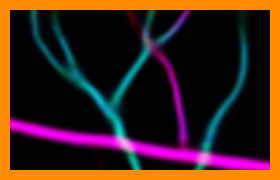}
    \includegraphics[height=0.065\textwidth]
        {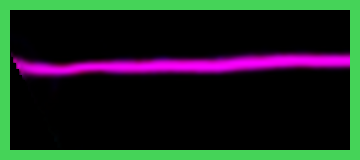}
    \includegraphics[height=0.065\textwidth]
        {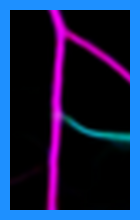}
    &
    \includegraphics[height=0.065\textwidth]
        {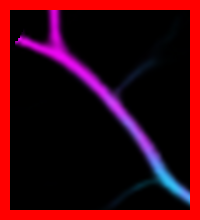}
    \includegraphics[height=0.065\textwidth]
        {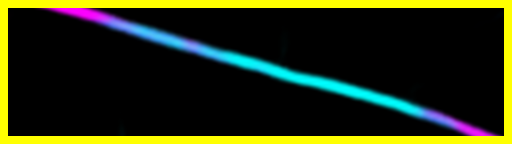}
    \includegraphics[height=0.065\textwidth]
        {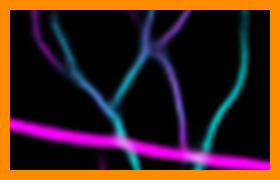}
    \includegraphics[height=0.065\textwidth]
        {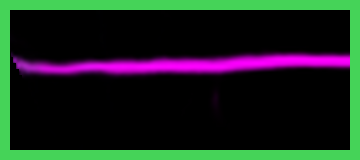}
    \includegraphics[height=0.065\textwidth]
        {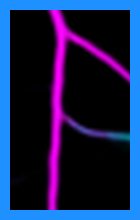}
    \\~\\

    \citet{Chen_MIA_2021} & \citet{galdran2022sota} & \citet{karlsson2022artery} \\

    \includegraphics[width=0.32\textwidth]
        {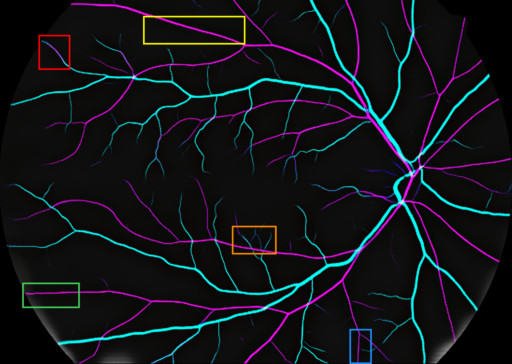}
    & \includegraphics[width=0.32\textwidth]
        {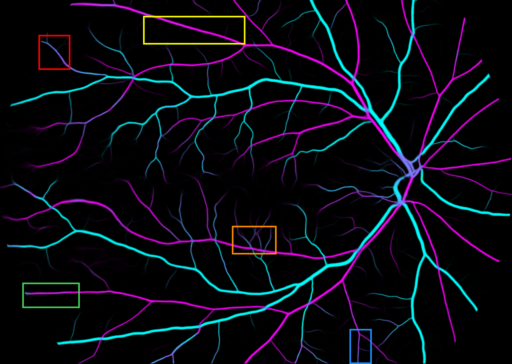}
    & \includegraphics[width=0.32\textwidth]
        {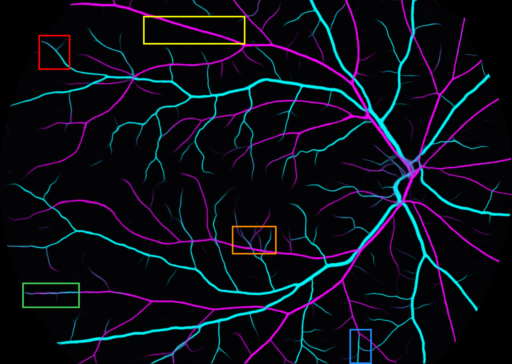}
    \\

    \includegraphics[height=0.065\textwidth]
        {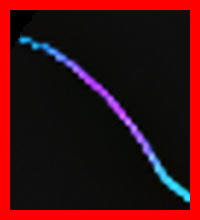}
    \includegraphics[height=0.065\textwidth]
        {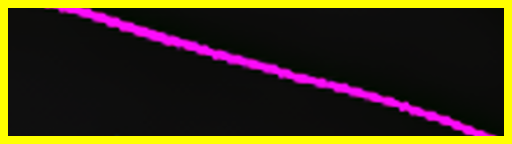}
    \includegraphics[height=0.065\textwidth]
        {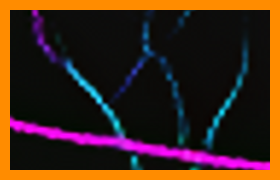}
    \includegraphics[height=0.065\textwidth]
        {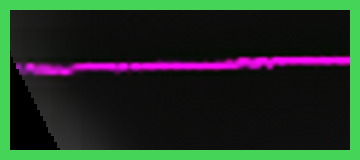}
    \includegraphics[height=0.065\textwidth]
        {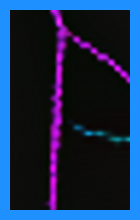}
    &
    \includegraphics[height=0.065\textwidth]
        {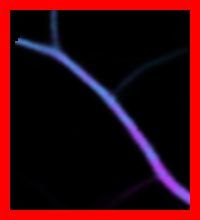}
    \includegraphics[height=0.065\textwidth]
        {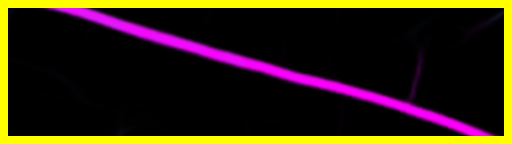}
    \includegraphics[height=0.065\textwidth]
        {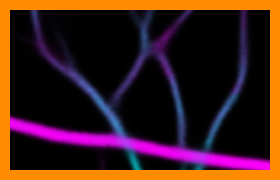}
    \includegraphics[height=0.065\textwidth]
        {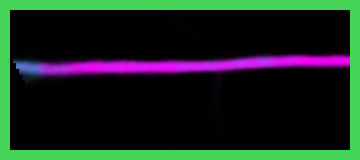}
    \includegraphics[height=0.065\textwidth]
        {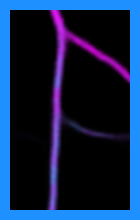}
    &
    \includegraphics[height=0.065\textwidth]
        {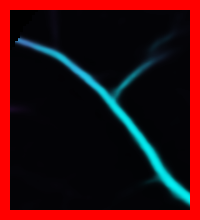}
    \includegraphics[height=0.065\textwidth]
        {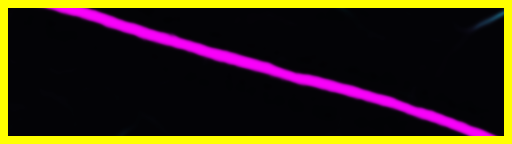}
    \includegraphics[height=0.065\textwidth]
        {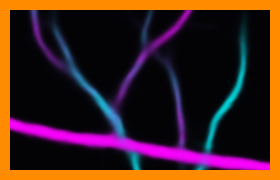}
    \includegraphics[height=0.065\textwidth]
        {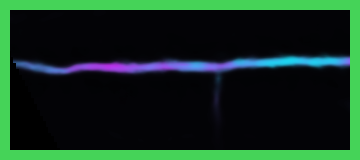}
    \includegraphics[height=0.065\textwidth]
        {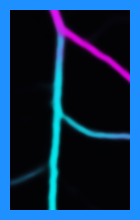} \\

    \end{tabularx} \\
    \caption{
        Examples of segmentation maps obtained by the different segmentation models in HRF dataset.
        A few notable differences between the segmentation maps of the different models are highlighted in colored boxes.
        [HRF-test, image \texttt{03\_g}]
    }%
    \label{fig:examples_sota_HRF_1}
\end{figure*}

Table~\ref{tab:av_segmentation} also emphasizes the potential of the RR module as a post-processing step to enhance the segmentation maps generated by other methods.
Of the 90 values in the table, 78 are improved when the RR module is used as a standalone post-processing step.
Moreover, in 14 out of 27 cases (3 datasets $\times$ 9 metrics), the combination of a state-of-the-art method with the RR module lead to the best overall performance (in the remaining 13 cases, the best performance is achieved by RRWNet).
These findings strongly suggest that the RR module serves as a robust and efficient post-processing approach, capable of significantly enhancing the performance of existing methods.

Examples of segmentation maps obtained by the different models compared in Table~\ref{tab:av_segmentation} are shown in Figs.~\ref{fig:examples_sota_RITE_1} and~\ref{fig:examples_sota_HRF_1}, for RITE and HRF datasets, respectively\footnote{All the segmentation maps obtained by our RRWNet model for RITE, HRF and LES-AV datasets will be publicly available at \url{https://github.com/j-morano/rrwnet}.}.
Overall, the segmentation maps generated by RRWNet are more accurate and topologically consistent than the segmentation maps obtained by the other methods.
For example, while the other methods tend to mix the classification of a vessel that is difficult to classify, the proposed RRWNet is able to correctly classify the whole vessel as either an artery or a vein (see Fig.~\ref{fig:examples_sota_RITE_1}, blue box).
In addition, RRWNet correctly classifies pixels at vessel crossings as belonging to both the artery and vein classes simultaneously (in the figures, represented by white pixels), while the other methods, except \citet{Chen_MIA_2021}, tend to classify them as only one of the two classes or leave them unclassified (with low probability for both classes), leading to discontinuities in the segmentation maps.
This RRWNet behavior inherently leads to more topologically consistent segmentation maps.

\begin{figure*}
    \centering
    \begin{tabularx}{\textwidth}{ZZZ}
    GT & \citet{galdran2022sota} & \citet{galdran2022sota} with RR \\

    \includegraphics[width=0.32\textwidth]
        {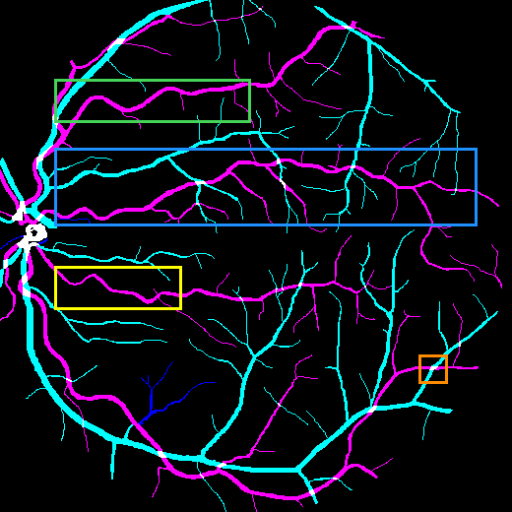}
    & \includegraphics[width=0.32\textwidth]
        {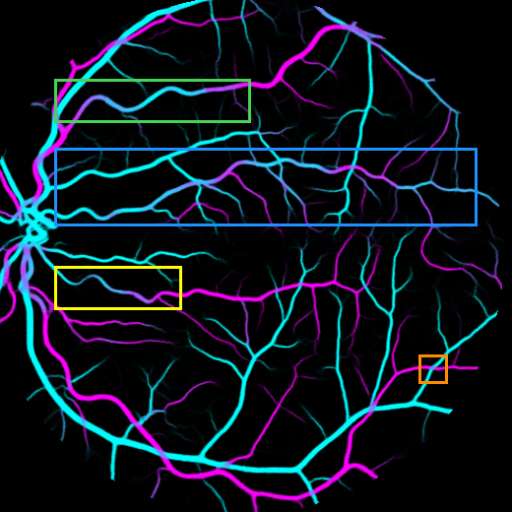}
    & \includegraphics[width=0.32\textwidth]
        {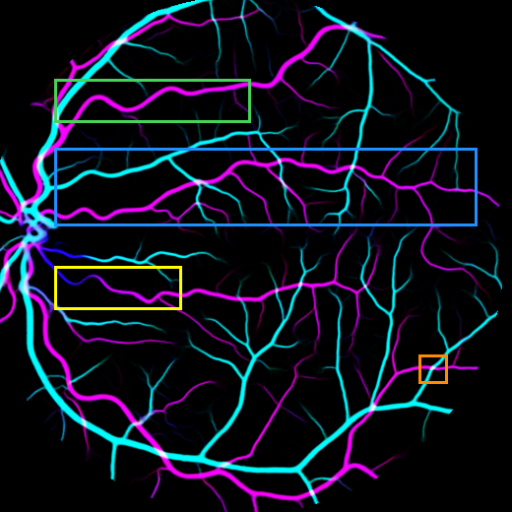}
    \\

    \includegraphics[height=0.05\textwidth]
        {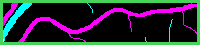}
    \includegraphics[height=0.06\textwidth]
        {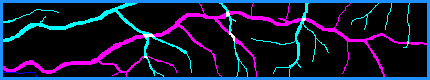}
    \includegraphics[height=0.07\textwidth]
        {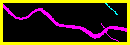}
    \includegraphics[height=0.07\textwidth]
        {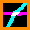}
    &
    \includegraphics[height=0.05\textwidth]
        {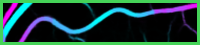}
    \includegraphics[height=0.06\textwidth]
        {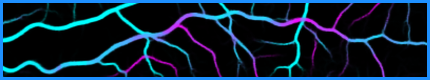}
    \includegraphics[height=0.07\textwidth]
        {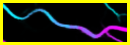}
    \includegraphics[height=0.07\textwidth]
        {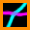}
    &
    \includegraphics[height=0.05\textwidth]
        {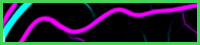}
    \includegraphics[height=0.06\textwidth]
        {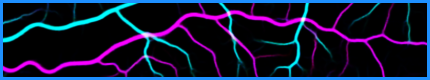}
    \includegraphics[height=0.07\textwidth]
        {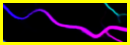}
    \includegraphics[height=0.07\textwidth]
        {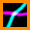}
    \end{tabularx} \\
    \caption{
        Examples of segmentation maps obtained by the model of \citet{galdran2022sota} before and after applying the proposed RR module as a post-processing step.
        A few notable differences between the resulting segmentation maps are highlighted in colored boxes.
        [RITE, image \texttt{11}]
    }
    \label{fig:examples_sota_rr_RITE}
\end{figure*}

Additionally, Fig.~\ref{fig:examples_sota_rr_RITE} shows examples of the segmentation maps obtained by the model proposed by \citet{galdran2022sota} before and after applying the proposed RR module as a post-processing step.
As evidenced by the figure, the use of this module  demonstrably enhances the quality of the segmentation results. Notably, the RR module effectively addresses the issue of false positive veins in the base segmentation map by accurately reclassifying them as arteries. Additionally, it successfully bridges the gaps between arteries and veins at crossing points (represented in white), where the original segmentation predicted a low probability for one or both classes (represented in dark purple).
As mentioned above, this issue is consistently observed in the outputs of other methods, with the exception of \cite{Chen_MIA_2021}.

The combined qualitative and quantitative evidence (as detailed in Table~\ref{tab:av_segmentation}) strongly suggests the efficacy of the proposed RR module as a generalizable post-processing step for improving the performance of diverse segmentation methods.

\section{Conclusions}

This work introduces RRWNet, a novel end-to-end deep learning framework specifically designed to address the challenge of manifest classification errors in semantic segmentation tasks, with a particular focus on A/V segmentation and classification.
These errors occur when the predicted segmentation violates the expected topological structure of the underlying object or structure being segmented.
To address this issue, RRWNet effectively combines stacking and recursive refinement approaches by decomposing the network into two specialized parts: a Base subnetwork for initial feature extraction and segmentation, and a Recursive Refinement subnetwork, which recursively refines the segmentation maps and iteratively resolves manifest classification errors.
With this design, RRWNet implicitly acknowledges the crucial role of both local and global features in achieving accurate segmentation.
The Base subnetwork utilizes local attributes like color and contrast, which FCNNs effectively capture, to generate initial segmentation maps.
However, for complex tasks like A/V segmentation, relying solely on local features is insufficient. To address this, the specialized Recursive Refinement subnetwork employs a recursive approach to capture and integrate global contextual information not readily apparent in local features. In addition, the iterative recursive process allows for gradual and significant refinement of the segmentation maps, leading to superior results compared to single-pass methods.
It is also important to note that this framework is not tied to a specific implementation, and is compatible with any FCNN architecture, so it can be easily integrated into existing FCNN-based methods.

To rigorously assess the efficacy of the proposed framework, we implemented a straightforward instantiation based on the well-established U-Net architecture. This implementation was evaluated on the task of A/V segmentation and classification within several publicly available retinography image datasets. The quantitative results demonstrated that the proposed method outperformed state-of-the-art methods by a notable margin, both in terms of A/V classification accuracy and, more remarkably, of topological consistency.
Furthermore, the standalone application of the proposed RR module demonstrably improved the segmentation maps generated by all the other compared methods, further substantiating its effectiveness as a generalizable post-processing step.

As a general framework, the proposed method has the potential to be applied to any semantic segmentation task where topological consistency plays a fundamental role in the segmentation quality.
Therefore, its application to other tasks, such as A/V segmentation in optical coherence tomography angiography (OCT-A) images and retinal layer segmentation in OCT images, represents a promising line of future work.

In conclusion, the proposed framework and its implementation represent an effective approach to A/V segmentation and classification, with the potential to be extended to other semantic segmentation tasks and modalities.
We believe that this work will serve as a good reference implementation and benchmark and encourage further research in this direction, and that it will contribute to the development of more robust and accurate semantic segmentation systems, with a particular focus on the field of ophthalmology.

\section*{CRediT authorship contribution statement}


José Morano: Conceptualization, Data curation, Formal analysis, Investigation, Methodology, Software, Validation, Visualization, Writing -- original draft, Writing -- review \& editing.
Guilherme Aresta: Conceptualization, Visualization, Writing -- review \& editing.
Hrvoje Bogunović: Funding acquisition, Project administration, Resources, Supervision, Writing -- review \& editing.

\section*{Declaration of Competing Interest}

The authors declare that they have no known competing financial interests or personal relationships that could have appeared to influence the work reported in this paper.

\section*{Acknowledgements}

This work was supported in part by the Christian Doppler Research Association, Austrian Federal Ministry for Digital and Economic Affairs, the National Foundation for Research, Technology and Development.


\bibliography{references}

\end{document}